\documentclass[aps,pra,floats,floatfix,twocolumn,longbibliography,superscriptaddress]{revtex4-2}
\usepackage{newtxtext}
\usepackage{amsmath}
\usepackage{amssymb}
\usepackage{color}
\usepackage{graphicx}
\usepackage{epsfig}
\usepackage{comment}
\usepackage{soul}
\usepackage{amsfonts}
\usepackage{mathrsfs}
\usepackage{mathtools}
\usepackage{float}
\usepackage{physics}
\usepackage{hyperref}
\hypersetup{colorlinks=true,
	citecolor=blue,
	linkcolor=blue,
	urlcolor=blue}

\begin{document}
\title{Quantum droplets and condensates in an optical lattice coupled to a dissipative cavity: Collective excitations and non-equilibrium dynamics}

\author{G. Vivek}
\affiliation{Indian Institute of Science Education and Research-Kolkata, Mohanpur, Nadia-741246, India}
\author{Sudip Sinha}
\affiliation{School of Physics, Korea Institute for Advanced Study, Seoul 02455, Korea}
\author{Subhasis Sinha}
\affiliation{Indian Institute of Science Education and Research-Kolkata, Mohanpur, Nadia-741246, India}

\begin{abstract}
Motivated by recent experiments on light-matter interacting systems, we investigate a dilute Bose gas and self-bound quantum droplets in a one-dimensional optical lattice coupled to a lossy cavity mode. Using a classical-field approach, we determine the stationary states and collective excitations of this non-equilibrium system. Apart from the usual Bogoliubov modes, we identify a polariton-like gapped excitation, the frequency of which softens as a precursor of the density ordering transition. Moreover, its relaxation time diverges as the critical point is approached, signaling the non-equilibrium nature of this transition. Dynamically, this polariton-like mode can be probed by inducing cavity field fluctuations, which in turn generates spatio-temporal oscillations of both the condensate and droplet states. In the droplet regime, we also analyze the bound modes which bear the characteristics of such non-equilibrium self-bound state. In addition, we uncover solitonic non-equilibrium states, including condensate with kink-like configuration and double-droplet state, and investigate their robustness following a sudden quench. Remarkably, although these states become unstable beyond a critical coupling, they continue to manifest in the dynamics, akin to the scarring phenomena. Our results demonstrate that dissipative cavity coupling provides a versatile route for exploring rich non-equilibrium dynamics of condensates and quantum droplets within experimentally accessible settings.
\end{abstract}
\date{\today}
\maketitle
\section{Introduction}
\label{sec:intro}
Open quantum many-body systems \cite{Petruccione_book, Weiss2008} have attracted considerable attention in recent years due to their potential to host a broad range of intriguing non-equilibrium phenomena, including relaxation dynamics, the emergence of exotic non-equilibrium phases and associated phase transitions, formation of continuous time crystals, dissipative chaos, and many others \cite{Daley2014, Weimer2021_RMP, Fazio2025, Zoller2008_Natphys,  Zoller2010_PRL, Parkins2007_PRA, Parkins2008_PRA, Kessler2012_PRA, Altman2013_PRL, Ciuti2013_PRL, Kulkarni2013_PRL, Carmichael2015_PRX, Koch2017_PRX, Casteels2017, Donner2018_PRL, Ciuti2018_PRA, Plenio2018_PRA, Agarwal2020_PRL, Millis2022_PRR, Kulkarni2022_PRA, Rakovszky2024_PRX, Nori2024_PRL, Sayak2025_PRL, Diehl2025_RMP, Fazio2018_btc_PRL, Hemmerich2021_PRL, Hemmerich2022_Science, Pohl2024_Nature, Pohl2025_PRL, Haake1988_PRL, Prosen2019_PRL, Denisov2019_PRL, Prosen2020_PRX, Piazza2015_PRL, Sayak2022_PRL, Santos2024_PRL, Mondal2025_PRL, Savona2025_PRR, Robnik2025_PRE, Mondal2025_PRE, Mondal2026_PRL}. Moreover, dissipation is inevitable in realistic physical systems, which raises important questions regarding the fate of an isolated quantum system \cite{Weiss2008}. In this context, atom-photon interacting systems inside optical cavities provide a natural platform for studying open quantum systems with tunable dissipation \cite{Daley2014, Esslinger2013_review, Ritsch_cavityQED_review, Keeling2018_PRX}. Such systems enable the exploration of fascinating non-equilibrium dynamical states that are otherwise difficult to realize in conventional solid-state materials \cite{Esslinger2010_Nat, Hemmerich2015_PNAS, Parkins2014_PRL, Parkins2017_PRL, Parkins2017_optica, Esslinger2016_Nat, Esslinger2017_Nat, Thomas2017_PRX, Bloch2017_PRL, Keeling2018_PRL, Ritsch2017, Rey2020_Nat, Esslinger2021_PRX, Carusotto2022_Nat_rev, Rey2024_Nat}.

A recent experiment demonstrated the emergence of a long-sought self-organized supersolid phase in cold atoms loaded into a two-dimensional optical lattice and coupled to a cavity mode, arising from cavity-induced long-range interactions \cite{Esslinger2016_Nat}. This opens up the possibility to study both non-equilibrium dynamics and correlated phases of quantum many-body systems in an open environment, where the photon loss from the cavity is the main source of dissipation. Apart from the transitions, the collective modes also play a crucial role for determining the dynamical stability of non-equilibrium phases. Moreover, tunability of the parameters in ultracold atomic setups provides the opportunity to realize various intriguing quantum many-body phases \cite{Dalibard2008_review, Bloch_review}. In this context, self-bound quantum droplets have emerged as a fascinating phase of quantum matter \cite{Petrov2015, Pfau2016, Tarruell2018, Modugno2018, Ferlaino2016, Pfau2021_review, Malomed2021_review}. These quantum droplets are stabilized by the competition between mean-field interactions and beyond mean-field quantum corrections \cite{Petrov2015, Pfau2021_review, Malomed2021_review, Astrakharchik2016, Astrakharchik2018, Malomed2020}. Unlike homogeneous condensates, such phases support localized collective excitations like breathing mode which reflects their self-bound structure \cite{Petrov2015, Astrakharchik2018, Malomed2020, Liu2020PRA, Sinha2023}. Furthermore, the dynamics of such self-bound macroscopic states have also attracted a lot of interest due to their non-linear behavior \cite{Astrakharchik2018, Modugno2019_PRL}.

It is therefore particularly interesting to explore the behavior of these droplets in the presence of single as well as multi-mode cavities \cite{Karpov2022, Ritsch2023_PRL, Pelster2025a, Pelster2025b, Parish2026}. In particular, understanding the fate of photon-coupled quantum droplets in a lossy cavity remains a pertinent issue. The presence of an optical lattice further enriches this physics by supporting discrete, or lattice, droplet phases \cite{Malomed2019, Zhong2019, Astrakharchik2020, Chen2021, Yamamoto2022, Boudjemaa2023, Susanto2024, Astrakharchik2024, Otajonov2025}. Coupling such lattice droplets to lossy cavity modes provides a natural route to realizing intrinsically non-equilibrium, self-bound states, in which light–matter interactions can drive collective structural transformations and emergent density ordering. This setting offers a promising platform for exploring the interplay between quantum fluctuations, lattice effects, and driven-dissipative dynamics. It also provides an opportunity to uncover novel collective behavior and excitations of self-bound droplets across non-equilibrium transitions induced by the cavity mediated interaction. In addition, beyond single-mode cavities, recent studies have shown that long-range interactions mediated by multi-mode cavities can themselves stabilize self-bound droplets \cite{Karpov2022}. These developments highlight the versatility of cavity Quantum Electrodynamics (QED) platforms for studying the collective behavior of condensates and self-bound droplets in dissipative environments, and for exploring the emergence of non-equilibrium phases and their associated dynamics.

In this work, we consider dilute bosons in a one-dimensional (1D) optical lattice coupled to a single leaky cavity mode. Within the mean-field approximation, we investigate the formation of a density-ordered supersolid phase driven by the atom-photon interaction in this non-equilibrium setting. Extending this analysis to the self-bound lattice droplet,  we find that a similar scenario of cavity mediated structural reorganization gives rise to a distinct superradiant droplet phase. In both the cases, we obtain the collective excitations and identify polariton-like modes which are strongly coupled to the photon field. The divergence in the relaxation time of such mode near the critical coupling strength provides a clear signature of the non-equilibrium nature of the density ordering transition. We further show that fluctuations of the cavity field can dynamically excite this mode, leading to characteristic spatio-temporal oscillations of the condensates and droplets. Finally, we uncover additional non-equilibrium configurations, including the double-droplet states, which can persist dynamically even beyond the regime of their stability.

\section{Model and Formalism}
\label{sec:formalism}
We consider weakly interacting bosons in a quasi one-dimensional optical lattice confined by a transverse harmonic trap of frequency $\omega_{\perp}$ with the corresponding trapping length $l_{\perp}=\sqrt{\hbar/m\omega_{\perp}}$. Within the mean-field limit and the tight-binding approximation, the energy functional of the bosons can be written as,
\begin{eqnarray}
	&&\mathcal{E}_{b}[\phi_{j}, \phi^{*}_{j}] =\notag\\
	&&\sum_{j}-J(\phi^{*}_{j}\phi_{j+1}+\phi^{*}_{j+1}\phi_{j}-2|\phi_{j}|^2) + V(|\phi_{j}|^2),
	\label{energy_functional_discrete}
\end{eqnarray}
where $\phi_{j}$ is the classical field at $j$th lattice site describing a condensate of $N$ bosons satisfying the normalization condition $\sum_{j}^{L} |\phi_{j}|^2 = N$, with $L$ being the number of lattice sites. The first term in the above equation describes the kinetic energy of the condensate with $J$ being the tunneling amplitude between the nearest neighbor sites. The term $2|\phi_{j}|^2$ is retained to recover the Laplacian in the continuum limit and corresponds to a uniform shift in the single-particle energy.
The second term describes a non-linear interaction potential $V$ which depends on the site density $n_{j} = |\phi_{j}|^2$, and is discussed later.
It should be noted that this classical-field approximation is valid only in the deep superfluid regime \cite{Pitaevskii2016_book} and the strongly correlated phases like Mott-Insulators are beyond the scope of the present approach.

Next, we consider that the system is coupled to a cavity mode, the energy functional for which can be written as,
\begin{eqnarray}
	\mathcal{E}[\phi_{j}, \phi^{*}_{j}, \alpha, \alpha^{*}]  &=& \mathcal{E}_{b}[\phi_{j}, \phi^{*}_{j}] + \hbar\omega_{0}|\alpha|^2 \notag\\
	&&-\sum^{L}_{j'} \frac{\lambda}{\sqrt{L}}(\alpha + \alpha^{*})(-1)^{j'}|\phi_{j'}|^2,
	\label{full_energy_functional_discrete2}
\end{eqnarray}
where $\alpha(t)$ is the amplitude of the cavity mode, $\omega_{0}$ is the corresponding frequency, and $\lambda$ denotes the coupling strength between the condensate and the cavity mode which favors a staggered density modulation.

Using the canonical EOMs, $\dot{\imath} \dot{\phi}_{j}(t) = \partial \mathcal{E}/\partial \phi^{*}_{j}$ and $\dot{\imath}\dot{\alpha}(t) = \partial \mathcal{E}/\partial \alpha^{*}$, we can write the discrete equations for the classical fields,
\begin{subequations}
	\begin{eqnarray}
		\dot{\imath}\dot{\phi}_{j} &=& -J(\phi_{j+1} + \phi_{j-1} - 2\phi_{j}) + V'(n_j)\phi_{j} \notag\\
		&&-\frac{\lambda}{\sqrt{L}}(\alpha + \alpha^{*})(-1)^{j}\phi_{j}, \label{condensate_field}\\
		\dot{\imath}\dot{\alpha} &=& \,\,\omega_0 \alpha - \frac{\lambda}{\sqrt{L}}\sum^{L}_{j'}(-1)^{j'}|\phi_{j'}|^2, \label{cavity_field}
	\end{eqnarray} \label{extended_GPE}
\end{subequations}
where $V'(n_{j}) = \frac{\partial V}{\partial n_{j}}$. The above equation of boson fields without the photon coupling is a generalization of the discrete non-linear Schr\"{o}dinger (DNLS) equation \cite{DNLS1988, Kevrekidis2009_book}.
In the continuum limit, these equations reduce to the extended Gross-Pitaevskii-equations \cite{Petrov2015, Astrakharchik2016}. Since the real cavities are lossy in nature, we can incorporate a  linear damping term in the equation Eq.~\eqref{cavity_field} for the cavity field, which is consistent with the mean-field limit of the Lindblad description \cite{Parkins2007_PRA, Keeling2010_PRL, Esslinger2013_review},
\begin{eqnarray}
	\dot{\imath}\dot{\alpha} &=& \,\,(\omega_0-\dot{\imath}\kappa) \alpha - \frac{\lambda}{\sqrt{L}}\sum^{L}_{j'}(-1)^{j'}|\phi_{j'}|^2, \label{cavity_field_dissipative}
\end{eqnarray}
where $\kappa$ denotes the photon decay rate. Incorporation of the photon loss term captures the feedback of the dissipative cavity on the condensate dynamics.

\subsection{Steady states}
We first discuss the dynamical steady states (stationary solutions) of the extended DNLS in Eq.~\eqref{extended_GPE}, for which the classical variables assume the following form,
\begin{eqnarray}
	\phi_j(t) = \bar{\chi}_j e^{-\dot{\imath} \mu t},\quad
	\alpha(t) = \bar{\alpha},
	\label{ss_eqns}
\end{eqnarray}
where $\bar{\chi}$, $\bar{\alpha}$ are time independent and $\mu$ denotes the chemical potential. Substituting Eq.~\eqref{ss_eqns} into Eqs.~\eqref{condensate_field}, \eqref{cavity_field_dissipative} yields the following steady state equations,
\begin{subequations}
	\begin{eqnarray}
		\mu \bar{\chi}_j &=& - J (\bar{\chi}_{j+1} + \bar{\chi}_{j-1} - 2\bar{\chi}_j) + V'(|\bar{\chi}_j|^2)\bar{\chi}_j \nonumber \\				&&- \frac{\lambda}{\sqrt{L}} (\bar{\alpha} + \bar{\alpha}^*) (-1)^j \bar{\chi}_j ,\label{SST1}\\
		\bar{\alpha} &=& \frac{\lambda}{(\omega_0 - \dot{\imath} \kappa)\sqrt{L}}
		\sum_{j'=1}^{L} (-1)^{j'} |\bar{\chi}_{j'}|^2. \label{SST2}
	\end{eqnarray} \label{SST}
\end{subequations}
It can be easily seen from the above equations that the steady state of the cavity field depends solely on the staggered density imbalance of the condensate, which we denote by $\mathcal{I}=\sum_{j=1}^{L} (-1)^{j} |\bar{\chi}_{j}|^2/L$, resulting in a self-consistently generated staggered potential acting on the condensate atoms. Within the adiabatic elimination approach, where we assume the cavity field relaxes faster than the atomic degrees of freedom, we can obtain an effective energy functional corresponding to the bosonic condensate by substituting Eq.~\eqref{SST2} into Eq.~\eqref{full_energy_functional_discrete2}, given by,
\begin{eqnarray}
	\mathcal{E}_{\rm eff}[\bar{\chi}_{j},\bar{\chi}^{*}_{j}] &=& \mathcal{E}_{b}[\bar{\chi}_{j},\bar{\chi}^{*}_{j}] \notag\\
	&&- \frac{\lambda^2 \omega_0}{(\omega_0^2 + \kappa^2)L}\!\left(\!\sum^{L}_{j'}(-1)^{j'}|\bar{\chi}_{j'}|^2\!\right)^{\!\!2}. \label{eff_energy}
\end{eqnarray}
Although the steady states characterizing the different phases can be obtained from the minimization of the above energy functional with appropriate boundary conditions, capturing the full non-equilibrium dynamics of the atom-photon coupled system is beyond the scope of this effective Hamiltonian.

\subsection{Stability analysis}
\label{subs:stability_analysis}
To analyze the dynamical stability of a steady state solution $\{\bar{\chi}_j,\bar{\alpha}\}$, we introduce small amplitude fluctuations around it,
\begin{equation}
	\chi_j(t)=\bar{\chi}_j+\delta\chi_j(t),
	\qquad
	\alpha(t)=\bar{\alpha}+\delta\alpha(t).
\end{equation}
Retaining only the terms linear in fluctuations in the full equations of motion (EOMs) [Eqs.~\eqref{condensate_field},\eqref{cavity_field_dissipative}], we obtain the EOMs for the fluctuations in the following form,

\begin{equation}
	\dot{\imath}\frac{d}{dt}
	\begin{pmatrix}
		\delta\chi_{j} \\
		\delta\chi_{j}^{*} \\
		\delta\alpha \\
		\delta\alpha^*
	\end{pmatrix}
	=
	\mathcal{J}
	\begin{pmatrix}
		\delta\chi_{j} \\
		\delta\chi_{j}^{*} \\
		\delta\alpha \\
		\delta\alpha^*
	\end{pmatrix},
	\label{eq:generalBdG}
\end{equation}
where $\mathcal{J}$ is the stability matrix evaluated at the steady state. In the absence of cavity coupling, diagonalizing the linearized equations recovers the conventional Bogoliubov excitation spectrum of the condensate. To determine the collective modes of the coupled atom-cavity system, we consider fluctuations of the form $\sim e^{\Lambda t}$ and solve the eigenvalue equation $|\mathcal{J}-\dot{\imath}\Lambda\mathcal{I}|=0$. The complex eigenvalues $\Lambda$, which appear in complex conjugate pairs, therefore characterize both the excitation spectrum and the dynamical stability of the steady state. Writing $\Lambda=\Gamma+\dot{\imath}\omega$, the real part $\Gamma=\mathrm{Re}(\Lambda)$ determines the damping or growth rate of the collective mode, while the imaginary part $\omega=|\mathrm{Im}(\Lambda)|$ gives its oscillation frequency. Dynamical stability requires all eigenvalues to satisfy $\Gamma\leq0$, such that fluctuations remain bounded or decay over time. Conversely, the onset of instability is signaled by at least one mode acquiring $\Gamma>0$, corresponding to exponentially growing fluctuations and breakdown of the steady state.

\section{Homogeneous Phase and Excitations}
\label{sec:homogeneous}
We first consider the homogeneous limit, where the bosons have an on-site interaction $u$,
\begin{eqnarray}
	V = \frac{u}{2}|\phi_{j}|^4. \label{mf_pot}
\end{eqnarray}
In the absence of atom-photon coupling, Eq.~\eqref{condensate_field} then reduces to the well-known DNLS equation, and the ground state therefore corresponds to a homogeneous condensate with equal occupation on every lattice site. Even for weak coupling $\lambda$, the stable steady state is described by,
\begin{eqnarray}
	|\bar{\chi}|^2 = \mu/u,\quad \bar{\alpha} = 0. \label{hom_st_below_critical}
\end{eqnarray}
Above a critical coupling strength $\lambda_{c}$, the aforementioned steady state in Eq.~\eqref{hom_st_below_critical} becomes unstable and a density modulated phase emerges.

\begin{figure}
	\centering
	\includegraphics[width=\columnwidth]{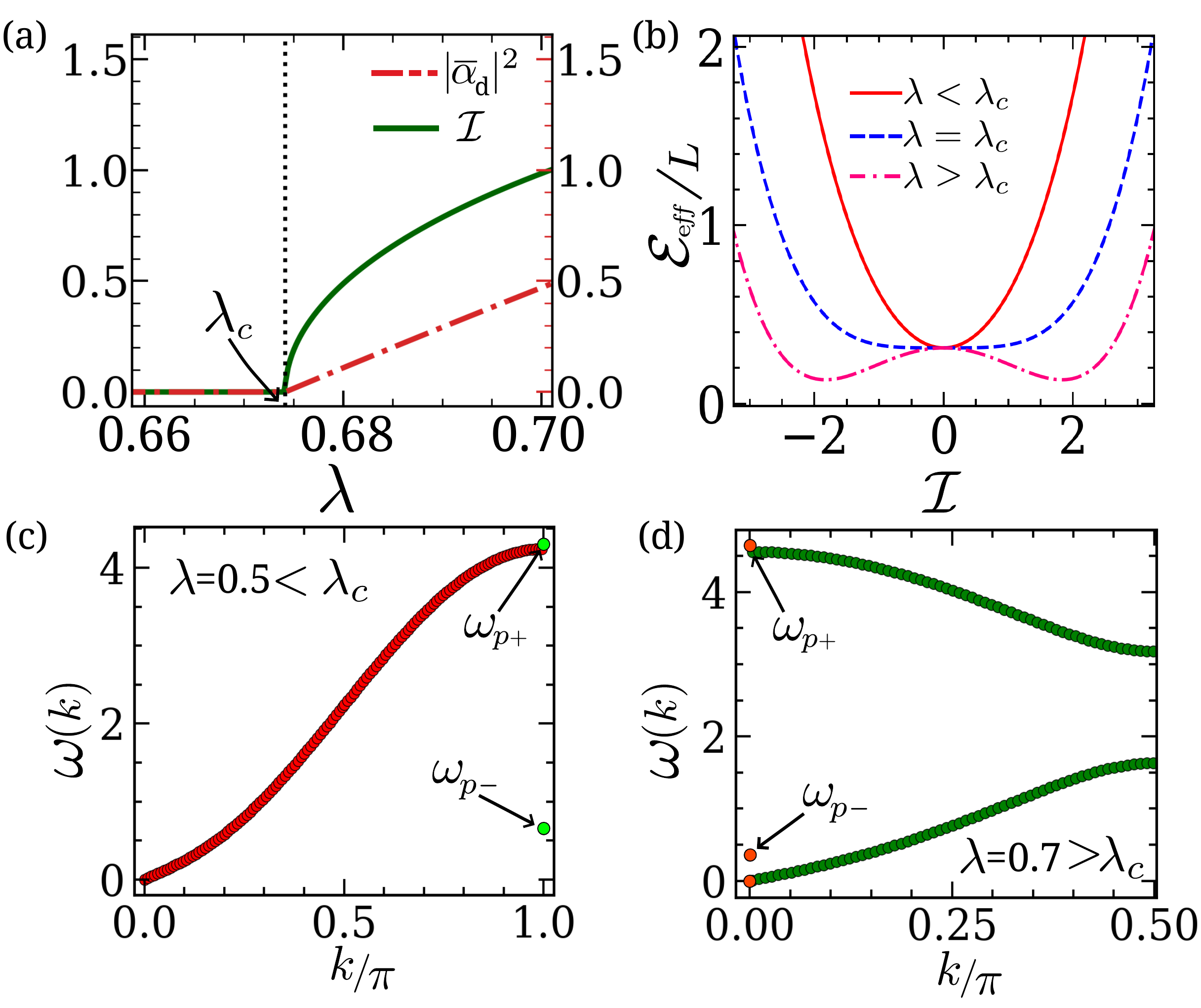}
	\caption{\textit{Excitations of the homogeneous condensate across the superradiant phase transition}: (a) The variation of the density imbalance $\mathcal{I}$ and the normalized photon number $|\bar{\alpha}_d|^2$ as a function of $\lambda$, across $\lambda_c$. (b) Effective energy functional in the Landau-Ginzburg form as a function of the order parameter $\mathcal{I}$, illustrating the $\mathbb{Z}_2$ symmetry-breaking transition across $\lambda_c$. Excitation spectrum of the respective phases (c) before and (d) after the transition, where the collective polariton-like mode in each case is highlighted with a different color. In this and the rest of the figures, we scale all the energies (times) by $J$ ($1/J$) and set $\omega_0=1$, unless otherwise mentioned. Parameters chosen: $n_0=2.5$, $u=0.1$, and $\kappa=0.1$.}
	\label{fig1}
\end{figure}

Notably, the coupled atom–cavity system possesses a discrete $\mathbb{Z}_2$ symmetry associated with the transformation \cite{Dicke_original, Emary2003_PRE, Baumann2011_PRL},
\begin{eqnarray}
	\phi_j \rightarrow \phi_{j+1},\quad \alpha \rightarrow -\alpha,
\end{eqnarray}
which corresponds to shifting the lattice to the left or right while simultaneously reversing the sign of the cavity field. Physically, breaking this symmetry corresponds to a super-radiant density modulation with wavevector $k=\pi$ and $|\alpha| \neq 0$. To analyze the formation of the density-modulated structure arising from the instability of the homogeneous states, we introduce a two-sublattice ansatz for which the condensate wavefunctions are parametrized by the density imbalance $\mathcal{I}$ between the even and odd lattice sites, given by,
\begin{eqnarray}
	\bar{\chi}_{\mathrm{even}} = \sqrt{n_0+\mathcal{I}},\quad
	\bar{\chi}_{\mathrm{odd}} = \sqrt{n_0-\mathcal{I}},
	\label{eq:two_sublattice_ansatz}
\end{eqnarray} 
where $n_{0}$ is the average on-site density. Within the adiabatic elimination approach, we can expand the effective energy functional corresponding to the Bose condensate [Eq.~\eqref{eff_energy}] in terms of $\mathcal{I}$ as follows,
\begin{eqnarray}
	\mathcal{E}_{\rm eff}[\mathcal{I}]/L &=& c + b\mathcal{I}^2 + a\mathcal{I}^4 + \mathcal{O}(\mathcal{I}^6), \label{LG_eq}
\end{eqnarray}
where $c=un_0^2/2$, $a=J/(4n_0^3)$, and the coefficient of the quadratic term is given by,
\begin{eqnarray}
	b &=& \frac{J}{n_0}+\frac{u}{2}-\frac{\lambda^2\omega_0}{\omega_0^2+\kappa^2}.
\end{eqnarray}
The last term in the above equation arises from the cavity-mediated interaction. Notably, near the transition, the energy functional given in Eq.~\eqref{LG_eq} is recast in the Landau-Ginzburg (LG) form, where the imbalance $\mathcal{I}$ serves as the order parameter. The invariance of the energy function for $\mathcal{I} \rightarrow -\mathcal{I}$ reflects the $\mathbb{Z}_{2}$ symmetry [see Fig.~\ref{fig1}(b)]. Consequently, the symmetry broken density-modulated phase (with $\mathcal{I} \neq 0$) emerges when the sign of $b$ changes ($b=0$), yielding the critical coupling strength,
\begin{eqnarray}
	\lambda_c &=& J\sqrt{\frac{\omega_{0}}{Jn_{0}}\left(1+\frac{\kappa^2}{\omega^2_{0}}\right)\left(1+\frac{un_{0}}{2J}\right)}.
\end{eqnarray}
Above the critical point ($b<0$), the order parameter, namely the Imbalance $\mathcal{I}=\pm\sqrt{-b/2a}$, increases smoothly from zero, signifying a continuous transition [see Fig.~\ref{fig1}(a)] where the superfluid phase changes to the density-modulated `supersolid' phase \cite{Esslinger2010_Nat, Recati2023_review, Sinha2025_review} with broken $\mathbb{Z}_{2}$ symmetry. In the vicinity of the critical point, the order parameter behaves as,
\begin{eqnarray}
	\mathcal{I}\sim(\lambda-\lambda_c)^{1/2}
\end{eqnarray}
for $\lambda>\lambda_c$, reflecting the mean-field nature of the transition \cite{Goldenfeld_book}. The density imbalance of the condensate also induces a non-vanishing photon occupancy $|\bar{\alpha}|^2\propto\mathcal{I}^2$ [see Eq.~\eqref{SST2}] above $\lambda_{c}$, similar to a continuous superradiant transition \cite{Dicke_original, Emary2003_PRE, Baumann2011_PRL}.  Consequently, the emergence of the cavity mediated density-ordered phase of the condensate is accompanied by the superradiant transition \cite{Esslinger2010_Nat, Esslinger2016_Nat}. The evolution of both the density imbalance and photon occupation across $\lambda_c$ is shown in Fig.~\ref{fig1}(a).

We now turn to the non-equilibrium steady states obtained directly from the coupled mean-field equations. Using a two-sublattice ansatz given in Eq.~\eqref{eq:two_sublattice_ansatz}, for which the even and odd lattice sites are allowed to acquire different densities, the steady state conditions given by Eqs.~\eqref{SST1} and \eqref{SST2} reduce to,
\begin{eqnarray}
	\mathcal{I} =
	\begin{cases}
		\hspace{2.cm}0\hspace{1.8cm}, & \lambda<\lambda_c, \\[1ex]
		\displaystyle
		\pm\sqrt{n_0^2-\left(\dfrac{2J}{2\lambda_{\rm eff} - u}\right)^2},
		& \lambda>\lambda_c ,
	\end{cases}
	\label{imbalance}
\end{eqnarray}
where $\lambda_{\rm eff}=\lambda^2\omega_0/(\omega_0^2+\kappa^2)$ denotes the effective cavity-mediated interaction strength. The steady state value of the photon field can be obtained by substituting the condensate density in Eq.~\eqref{SST2} and is given by,
\begin{eqnarray}
	\bar{\alpha}
	=
	\begin{cases}
		\hspace{1.5cm}0\hspace{0.9cm},
		& \lambda<\lambda_c, \\[1ex]
		\displaystyle
		\pm
		\sqrt{L}\,
		\frac{\lambda}{\omega_0-\dot{\imath}\kappa}\,
		\mathcal{I}\hspace{0.2cm},
		& \lambda>\lambda_c .
	\end{cases}
	\label{alpha_ss}
\end{eqnarray}
Since the photon field scales with system size, it is convenient to define the intensive quantity $\bar{\alpha}_{d}=\bar{\alpha}/\sqrt{L}$. The corresponding chemical potentials are given by, $\mu=un_0$ in the homogeneous phase and $\mu=2n_0(u-\lambda_{\rm eff})+2J$ in the staggered phase. Together, Eq.~\eqref{imbalance} and the corresponding chemical potentials determine the steady state configurations of the condensate. For $\lambda<\lambda_c$, the system remains in a homogeneous state with $\mathcal{I}=0$ and vanishing cavity field $\bar{\alpha}=0$, preserving the underlying $\mathbb{Z}_2$ symmetry. Above the critical coupling, two degenerate solutions with $\pm\mathcal{I}$ emerge, corresponding to symmetry-broken staggered density configurations with unequal occupations on the even and odd lattice sites. This density modulation is accompanied by the onset of a finite cavity field, $\bar{\alpha}\neq0$, signaling the superradiant phase of the coupled atom-cavity system. In the presence of dissipation, the cavity field relaxes toward a steady state value determined self-consistently from the atomic density modulation, as follows from Eq.~\eqref{SST2}. Consequently, during the dynamics, the coupled atom-photon system evolves toward one of the two symmetry-related superradiant steady states, as discussed later in Sec.~\ref{sec:real time and quench dynamics}.

\subsection*{Excitation Spectrum}
\label{sub:homogeneous_excitations}
To investigate both the dynamical stability of the non-equilibrium steady states and their excitation spectra, we perform a linear stability analysis of small-amplitude fluctuations around the stationary solutions, as outlined in Sec.II B. In the absence of the photon field, this reduces to the usual Bogoliubov theory, which yields the excitations of the stationary condensate at equilibrium. In the present framework, we retain fluctuations of the dissipative cavity field, which hybridize with the condensate modes and give rise to polariton-like excitations in addition to the conventional Bogoliubov modes. While the Bogoliubov excitation spectra carry the characteristic signatures of the condensate in the homogeneous and density-ordered phases, the decay rates of the polaritonic excitations exhibit critical slowing down in the vicinity of the transition, signaling its non-equilibrium nature. For comparison, we also calculate the Bogoliubov excitations of the effective Hamiltonian obtained by adiabatic elimination of the photon mode, which is unable to capture such polaritonic modes. Although the effective Hamiltonian can reproduce the critical coupling strength and the stationary phases, the signatures of the non-equilibrium transition encoded in the excitation spectrum are beyond its scope. In the following, we calculate the excitation spectrum in detail separately above and below the density-ordering transition.

The excitation spectrum not only reflects the properties of both the homogeneous and density-modulated configurations, owing to its dependence on the structure of the stationary state, but also carries a distinct signature of the non-equilibrium transition. In the uniform phase, the condensate preserves the discrete translational symmetry of the lattice, whereas above the transition it develops a two-sublattice density-wave structure. To describe both phases within a common framework, we represent the steady-state wavefunction of the condensate at site $j$ as,
\begin{equation}
	\bar{\chi}_j=\bar{\chi}+\epsilon (-1)^j,
	\label{steady_state_epsilon}
\end{equation}
where $\bar{\chi}$ and $\epsilon$ denote the average condensate amplitude and its modulation, respectively. For a fixed average number of particles per site $n_0$, the density modulation can be written as $\mathcal{I}=2\bar{\chi}\epsilon$ and $\bar{\chi}^2=n_0-\epsilon^2$.
The condensate fluctuations are decomposed into Fourier modes as $\delta\chi_j(t)=L^{-1/2}\sum_k\delta\chi_k(t)e^{ikj}$, where $k$ belongs to the first Brillouin zone, $-\pi\leq k\leq\pi$. Here, momenta are measured in units of the inverse lattice spacing, (1/a). In momentum space, the linearized EOMs of the fluctuations [given in Eq.~\eqref{eq:generalBdG}] take the form,
\begin{subequations}
	\begin{eqnarray}\label{eq:momentum_fluctuations}
		\dot{\imath}\delta\dot{\chi}_k&=&\big[2J(1-\cos(k))+2un_0-\mu\big]\delta\chi_k+un_0\delta\chi_{-k}^* \nonumber\\
		&+&\big[4u\bar{\chi}\epsilon-\lambda\left(\alpha_{d}+\alpha_{d}^{*}\right) \big]\delta\chi_{k-\pi}+2u\bar{\chi}\epsilon\delta\chi_{\pi-k}^* \nonumber\\
		&-&\lambda\big[\bar{\chi}\delta_{k,\pi}+\epsilon\delta_{k,0}\big]\left(\delta\alpha+\delta\alpha^*\right)\\
		\dot{\imath}\delta\dot{\alpha}&=&\left(\omega_0-\dot{\imath}\kappa\right)\delta\alpha-\lambda\bar{\chi}\left(\delta\chi_{\pi}+\delta\chi_{\pi}^*\right)\nonumber \\
		&&\hspace{1.95cm}-\lambda\epsilon\left(\delta\chi_0+\delta\chi_0^*\right). \label{eq:photon_k_fluctuations}
	\end{eqnarray}
\end{subequations}
Similar equations can also be written for the modes $\delta\chi_{-k}^*$ and $\delta\alpha^*$. Assuming that the time evolution is of the form $\sim e^{\Lambda t}$, the equations of fluctuation modes $(\delta\chi_k,\delta\chi_{-k}^*,\delta\chi_{k-\pi},\delta\chi_{\pi-k}^*,\delta\alpha,\delta\alpha^*)$ reduce to an eigenvalue problem where the eigenvalues $\Lambda_k$ are obtained by diagonalizing the $6\times6$ stability matrix. The steady state is dynamically stable if $\mathrm{Re}(\Lambda_k)\leq0$ for all eigenvalues and all momenta $k$ in the Brillouin zone. The excitation frequencies are given by $\omega_k=\mathrm{Im}(\Lambda_k)$, while $\mathrm{Re}(\Lambda_k)$ determine the corresponding damping rates. As evident from Eq.~\eqref{eq:momentum_fluctuations}, the photonic fluctuations couple only to the condensate fluctuation modes $\delta\chi_0$ and $\delta\chi_\pi$, corresponding to momenta $k=0$ and $k=\pi$, respectively. This hybridization between the Bogoliubov quasiparticles and photonic fluctuations gives rise to polariton-like collective excitations. In contrast, the remaining momentum modes are decoupled from the cavity-field fluctuations and correspond to conventional Bogoliubov excitations of the condensate with $\mathrm{Re}(\Lambda_k)=0$. Such hybridization is absent in the effective Hamiltonian description, which supports only Bogoliubov-like modes. Their detailed structure is described in Appendix~\ref{appendixB}. In the following, we discuss the collective excitations of the homogeneous and density-wave phases, separately.

\begin{figure}
	\centering
	\includegraphics[width=\columnwidth]{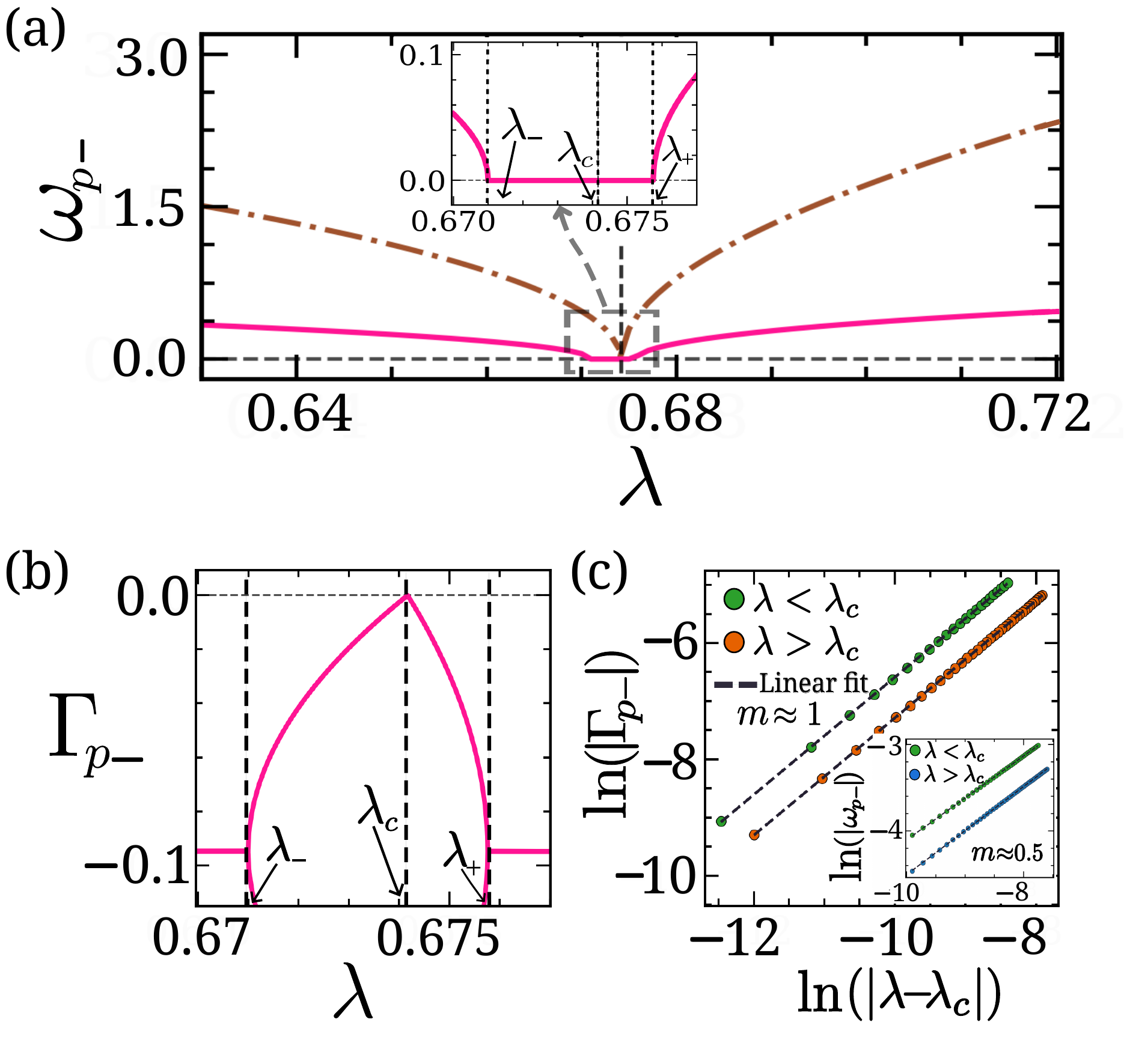}
	\caption{\textit{Signatures of dissipative phase transition in the homogeneous phase}: (a) Variation of frequency of the lowest collective mode $(\omega_{p-})$ with atom-photon coupling $\lambda$ across the transition point. This mode vanishes at a point $\lambda_-<\lambda_{c}$, while it becomes gapped at a coupling strength $\lambda_+>\lambda_c$ (see inset). The brown dashed-dotted line corresponds to the soft modes below and above the transition obtained using the effective Hamiltonian, the analytical forms of which are given in Appendix.~\ref{appendixB}. (b) Variation of the real part $\Gamma_{p-}$ of the eigenvalues corresponding to the lowest polaritonic branch across $\lambda_c$, which exhibits a bifurcation at $\lambda_-$ ($\lambda_+$) below (above) $\lambda_c$, accompanied by a gap closing at $\lambda_c$. (c) Scaling behavior of the real part (imaginary part in the inset) of the  eigenvalues corresponding to the lowest polaritonic branch across the transition (at the gap closing point $\lambda_\pm$). Parameters chosen: $n_0=2.5$, $u=0.1$, and $\kappa=0.1$.}
	\label{fig2}
\end{figure}

\textit{Homogeneous phase} $(\lambda<\lambda_c)$: In this regime, the steady state is described by a uniform condensate wavefunction with $\epsilon=0$ and a vanishing cavity field, $\bar{\alpha}=0$. For all condensate modes $(\delta\chi_k,\delta\chi_{-k}^{*})$ with $k\neq\pi$, the condensate and photon fluctuations remain decoupled. The corresponding $2\times2$ stability matrix yields purely imaginary eigenvalues $\Lambda=\pm\dot{\imath}\omega(k)$, with the collective excitation spectrum given by the usual Bogoliubov dispersion \cite{Pitaevskii2016_book},
\begin{equation}
	\omega(k)=\sqrt{
		\big[2J(1-\cos k)+un_0\big]^2-u^2n_0^2}.
	\label{eq:uni_continuum}
\end{equation}
In the long-wavelength limit, this gives a gapless sound mode $\omega\approx v_s|k|$ [see Fig.~\ref{fig1}(c)], with sound velocity $v_s=\sqrt{2Jun_0}$.

On the other hand, the condensate mode corresponding to $k=\pi$ hybridizes with the photon fluctuations, giving rise to a $4\times4$ stability matrix for the fluctuation vector $(\delta\chi_{\pi},\,\delta\chi_{-\pi}^{*},\,\delta\alpha,\,\delta\alpha^{*})$. The uniform condensate is stable when all eigenvalues satisfy $\mathrm{Re}(\Lambda)\leq0$, while their imaginary parts determine the corresponding excitation frequencies. The hybridization between the condensate and photon modes thus gives rise to two gapped polariton-like collective modes. In the absence of dissipation $(\kappa=0)$, the eigenvalues are purely imaginary, $\Lambda=\pm\dot{\imath}\omega_{\pm}$, and the frequencies of these two modes can be written as,
\begin{equation}
	\omega_{p\pm}^2
	=
	\frac{\big(\omega^2(\pi)+\omega_0^2\big)}{2}
	\pm
	\frac{1}{2}
	\sqrt{\big(\omega^2(\pi)-\omega_0^2\big)^2+64J\omega_0n_0\lambda^2},
	\label{eq:uni_polaritonic_modes}
\end{equation}
where $\omega(\pi)$ is the Bogoliubov excitation [given in Eq.~\eqref{eq:uni_continuum}] of the uncoupled condensate with momentum $k=\pi$. It is evident from Eq.~\eqref{eq:uni_polaritonic_modes}, for $\lambda\rightarrow0$, the lower and upper branches approach the photon ($\omega_{p-}\rightarrow\omega_0$) and Bogoliubov frequencies   ($\omega_{p+}\rightarrow\omega(\pi)$), respectively. These polaritonic modes closely resemble the familiar exciton-polariton branches \cite{Wouters2007_2009,Deng2010_review}, particularly near the resonance ($\omega(\pi)\sim\omega_0$) where an avoided level-crossing occurs due to their hybridization, leading to a splitting of the order $\lambda$. The lower polaritonic mode softens with increasing $\lambda$ and vanishes at the critical point, signaling the $\mathbb{Z}_2$ symmetry-breaking transition. With dissipation, the frequency of the lower polaritonic mode behaves similarly but picks up a finite damping rate $\Gamma_{p-}$, which are depicted in Fig.~\ref{fig2}(a,b) respectively. Notably, as non-zero $\kappa$, the frequency $\omega_{p-}$ vanishes before the critical coupling ($\lambda_-<\lambda_c$), the details of which will be discussed later in this section. Typically in bosonic systems with long-range interactions, the softening of the Bogoliubov mode at finite momentum is a precursor to the emergence of a density-modulated supersolid phase \cite{Mottl2012, Chomaz2018, Chomaz2019, Recati2023_review, Sinha2025_review}. In the present case, the softening of the Bogoliubov mode with $k=\pi$ is also induced by the cavity mediated long range interaction, signaling the onset of density-wave order with ordering wave vector $k=\pi$. We next analyze the excitations of the density-wave superradiant phase.

\textit{Density-modulated superradiant phase} $(\lambda>\lambda_c)$: In this regime, the steady state is characterized by a finite amplitude modulation of the wavefunction with $\epsilon\neq0$ and non-vanishing cavity field $\bar{\alpha}\neq0$. As follows from Eq.~\eqref{eq:momentum_fluctuations}, the photon fluctuations couple only to the $k=0$ and $k=\pi$ sectors of the condensate fluctuations, while all other momentum sectors remain decoupled from the cavity field. Furthermore, the two-sublattice structure of the condensate, characterized by $\epsilon\neq0$, couples the condensate fluctuations $\delta\chi_k$ and $\delta\chi_{k-\pi}$. Consequently, for $k\neq0,\pi$, the corresponding collective frequencies can be obtained by diagonalizing the $4\times4$ stability matrix associated with the fluctuations $(\delta\chi_{k},\,\delta\chi_{-k}^{*},\,\delta\chi_{k-\pi},\,\delta\chi_{\pi-k}^{*})$. Similar to the previous case, the eigenvalues remain purely imaginary $\Lambda(k)=\pm\dot{\imath}\omega(k)$, and provide the dispersion corresponding to the two branches of excitations as shown in Fig.~\ref{fig1}(d), which are given by,
\begin{eqnarray}
	\omega_{\pm}(k)^2
	&=& 4J^2\cos^2(k) + \mathcal{Q} \nonumber\\
	&&\pm \sqrt{
		\left(4J^2\cos^2(k)+\mathcal{Q}\right)^2-\mathcal{P}_k},
	\label{stag_continuum}
\end{eqnarray}
where $\mathcal{Q}$ and $\mathcal{P}_k$ are written as,
\begin{eqnarray}
	\mathcal{Q} &=& n_0^2[(4\lambda_{\mathrm{eff}}^2-u^2)(1+\Delta^2)+2\lambda_{\mathrm{eff}}u\Delta^2] \label{bog_details1}\\
	\mathcal{P}_k &=& n_0^4\{[4\lambda_{\mathrm{eff}}^2-u^2\!+\!(4\tilde{u}^2-u^2)\Delta^2]^2 \nonumber\\
	&&-\big[8\lambda_{\mathrm{eff}}\tilde{u}-2u^2\big]^2\!\Delta^2\}\!+\! 16J^4\cos^4(k)\nonumber\\
	&&- 8J^2n_0^2\cos^2(k)[4\lambda_{\mathrm{eff}}^2\!+\!u^2\!-\!(4\tilde{u}^2\!+\!u^2)\Delta^2], \label{bog_details2}
\end{eqnarray}
with $\tilde{u} = u-\lambda_{\mathrm{eff}}$ and $\Delta=\mathcal{I}/n_0$. Note that the two branches of the excitations reflect the two-sublattice structure of the steady state, for which the dispersion is defined in the reduced Brillouin zone $-\pi/2\leq k\leq\pi/2$. At the zone boundary, $|k|=\pi/2$, the two excitation branches are separated by a finite gap $\Delta\omega=\omega_+(\pi/2)-\omega_-(\pi/2)\approx\Delta$, just above the transition [see Fig.~\ref{fig1}(d)]. The opening of this gap is a direct consequence of the density modulation and provides a characteristic signature of the $\mathbb{Z}_2$ symmetry-broken density wave steady state.
Unlike the homogeneous phase, the dissipation rate $\kappa$ enters in the Bogoliubov frequencies [see Eqs.~\eqref{bog_details1} and \eqref{bog_details2}] through the effective coupling strength $\lambda_{\rm eff}$, however, their damping remains absent.

In this phase, the polaritonic excitations arise due to the mixing of the photon fluctuation with the condensate fluctuations both at $k=0$ and $k=\pi$ as seen from Eqs.~\eqref{eq:momentum_fluctuations} and \eqref{eq:photon_k_fluctuations}.
Diagonalizing the $6\times6$ stability matrix associated with the fluctuations $(\delta\chi_{0},\,\delta\chi_{0}^{*},\,\delta\chi_{-\pi},\,\delta\chi_{\pi}^{*},\,\delta\alpha,\,\delta\alpha^{*})$ yields two gapped polaritonic modes $\omega_{p\pm}$, along with a gapless Goldstone mode  $\omega_G=0$. The latter is the $k\rightarrow0$ continuation of the sound branch $\omega_{-}(k)$ ($\sim |k|$ for $k\ll 1$) given in Eq.~\eqref{stag_continuum}. The damping of the polaritonic modes with $\Gamma_{p\pm} \leq0$ also ensures the stability of the density-ordered supersolid phase. Both the frequency and the damping rate of the lowest polaritonic branch are shown in Fig.~\ref{fig2}(a,b), respectively.

We now take a closer look at the behavior of the polaritonic excitations at the vicinity of the critical point, which reflects the nature of the transition. Near the transition, the lower polaritonic branch $\omega_{p-}$ softens continuously from both sides of the critical point $\lambda_c$, and vanishes within the interval $\lambda_-\leq\lambda\leq\lambda_+$ surrounding $\lambda_c$, as shown in Fig.~\ref{fig2}(a). In particular, close to the couplings $\lambda_\pm$, $\omega_{p-}$ vanishes as $\sim\sqrt{|\lambda-\lambda_{\pm}|}$ (see Fig.~\ref{fig2}(c)). Notably, the size of this interval depends on the dissipation strength $\kappa$, and both $\lambda_\pm$ converge to $\lambda_c$ as $\kappa\rightarrow0$. For $\kappa\neq0$, within this interval, the real part of the eigenvalue $\Gamma_{p-}$ corresponding to the lower polaritonic mode, splits into two branches, the largest of which $\Gamma_{p-}^{m}$ vanishes at $\lambda_c$ as $\sim|\lambda-\lambda_c|$ from both sides [see Figs.~\ref{fig2}(b),(c)]. Consequently, the divergence of the relaxation time $1/\Gamma_{p-}^m$ near the critical point as $\sim 1/|\lambda-\lambda_c|$, signals the critical slowing down, unveiling the non-equilibrium nature of the transition \cite{Parkins2007_PRA, Parkins2008_PRA, Casteels2017}. Interestingly, the open Dicke model also exhibits a similar phenomenon at the non-equilibrium transition \cite{Parkins2007_PRA}. Dynamically, these polaritonic modes can be probed in a condensate by inducing fluctuations in the cavity field, which we discuss in Sec.~\ref{sec:real time and quench dynamics}. On the contrary, in the isolated system, the vanishing of the low lying mode occurs exactly at the critical point $\lambda_c$, corresponding to the QPT associated with the $\mathbb{Z}_2$ symmetry breaking. Such behavior can be recovered from the present system in the limit $\kappa\rightarrow0$, where the polaritonic mode $\omega_{p-}$ vanishes exactly at the critical point $\lambda_c$, with vanishing damping rate.

Within the adiabatic elimination approach, the effective Hamiltonian obtained after eliminating the photon field [see Eq.~\eqref{eff_energy}] yields the usual Bogoliubov excitations [see Eqs.~\eqref{eq:uni_continuum} and \eqref{stag_continuum}] except for $k=\pi$ below the transition and $k=0,\pi$ above the transition. The details of these excitations obtained using the same approach are discussed in Appendix.~\ref{appendixB}.
Such an effective description captures the equilibrium transition to the density-ordered phase and the softening of the low-lying modes. Notably, the gap of such low lying modes vanishes at the critical point [see Fig.~\ref{fig2}(a)], even in the presence of dissipation, similar to the $\mathbb{Z}_2$ symmetry breaking transition in the isolated system, contrasting the behavior of the polaritonic mode $\omega_{p-}$. Moreover, the frequency of the polaritonic mode $\omega_{p-}$ deviates from that of the low lying excitations obtained from the effective Hamiltonian, even in the absence of dissipation. Consequently, the manifestation of the non-equilibrium nature of the density-ordering transition through the polaritonic mode is beyond the scope of the effective Hamiltonian picture, which is primarily due to the absence of the photon fluctuations. In the next section, we perform the same analysis for the droplet phase to analyze its excitations in details.

\section{Droplet Formation}
\label{sec:droplet}
Next, we investigate non-equilibrium phenomena associated with quantum droplets formed in a binary mixture of Bose condensates loaded into a one-dimensional lattice and coupled to a dissipative cavity mode. Such self-bound droplets emerge from the competition between mean-field interactions and beyond-mean-field quantum fluctuations, commonly described by the Lee–Huang–Yang (LHY) correction \cite{Petrov2015,Pfau2021_review}.
For a symmetric binary Bose mixture with equal densities and identical intra-species interactions, the droplet phase can be effectively described by a single-component classical field obeying an extended Gross–Pitaevskii equation that includes the LHY correction term \cite{Astrakharchik2016, Astrakharchik2018, Malomed2020}.
\begin{figure}
	\centering
	\includegraphics[width=\columnwidth]{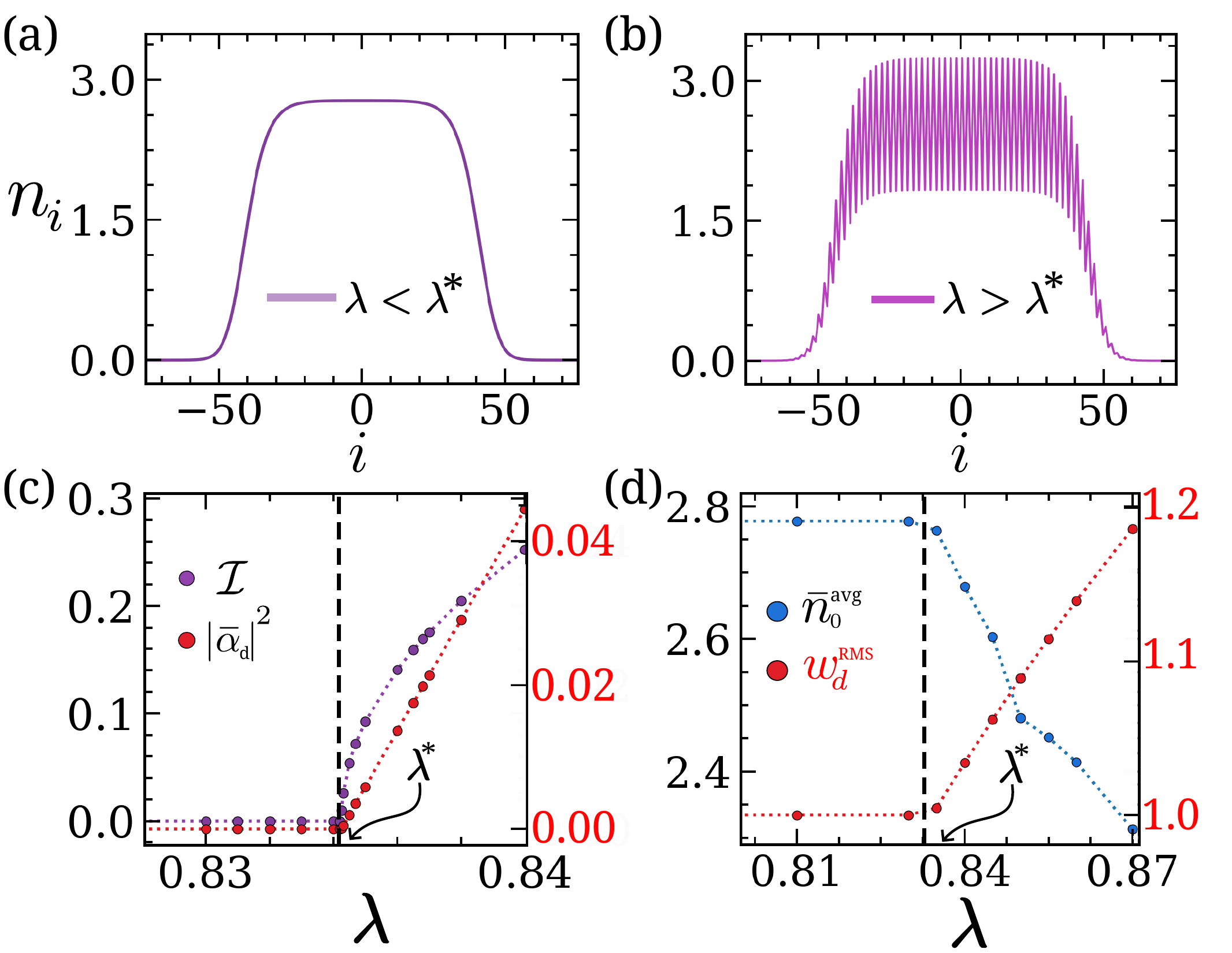}
	\caption{\textit{Properties of the steady states in the droplet regime ($\zeta\neq0$) across the crossover coupling $\lambda^*$}: (a) Flat-top droplet solution for $\lambda=0.5<\lambda^*$, characterized by an almost uniform density in the bulk. (b) Staggered (superradiant) droplet solution for $\lambda=0.85>\lambda^*$, where a density modulation is induced between the neighboring lattice sites. (c) Variation of the steady-state density imbalance $\mathcal{I}$ and the photon occupation density in the cavity $|\bar{\alpha}_d|^2$, across the crossover point. (d) The average central density $\bar{n}_0^{\rm avg}$ (blue circles, left axis) and the root mean square (RMS) width $w_d^{\rm RMS}$ (red circles, right axis) of the droplet have been plotted as a function of $\lambda$ across the crossover point. The RMS width has been normalized by the value corresponding to the flat-top droplet, which remains constant below the crossover. The crossover point $\lambda^* \simeq 0.834$ is marked by vertical dashed lines in (c,d). The discrete density distributions depicted here and in the rest of the manuscript have been obtained using periodic boundary conditions unless otherwise mentioned. The dotted lines in (c,d) only act as a guide to eye for the variation of the respective quantities. Parameters chosen: $n_{0}=1.47$, $u = 0.04$, $\zeta=0.1$, and $\kappa=0.1$.}
	\label{fig3}
\end{figure}
In a lattice, the on-site interaction potential can be written as,
\begin{eqnarray}
	V = \frac{u}{2}|\phi_{j}|^4 - \frac{2}{3}\zeta|\phi_{j}|^3, \label{lhy_pot}
\end{eqnarray}
where $\phi_{j}$ is the wavefunction at the $j$th site. The second term of on-site energy describes LHY correction with strength $\zeta$ [see also appendix.~\ref{appendixA}]. It should be noted that the LHY correction in the above equation is attractive in 1D, which counterbalances the repulsive non-linear on-site interaction, leading to the emergence of a stable self-bound quantum droplet phase \cite{Astrakharchik2016, Astrakharchik2018}.
We obtain the droplet state by minimizing the effective energy functional Eq.~\eqref{eff_energy} of the condensate after adiabatically eliminating the photon field. The numerical minimization is carried out using the imaginary-time propagation method with periodic boundary condition, yielding the ground-state wavefunction $\phi_j$ and the corresponding chemical potential $\mu$. The corresponding photon field can be obtained from the density imbalance using Eq.~\eqref{SST2}. Notably, these solutions also represent the non-equilibrium steady states of the dynamical equations [Eqs.~\eqref{condensate_field},\eqref{cavity_field_dissipative}].
In the regime of weak coupling to the photon field (small $\lambda$),
and for sufficiently large number of particles, the droplet corresponds to a flat-top density structure [see Fig.~\ref{fig3}(a)], for which the average on-site density and the corresponding chemical potential are respectively given by,
\begin{eqnarray}
	\bar{n}_{0} = \frac{4}{9}\left(\frac{\zeta}{u}\right)^2,\quad \mu = -\frac{2}{9}\frac{\zeta^2}{u}.\label{eq:droplet_sat}
\end{eqnarray}
Note that, apart from the particle number, the width of the droplet can also be controlled by the relative strength of the mean-field repulsion and the LHY term.

\begin{figure*}
	\centering
	\includegraphics[width=\textwidth]{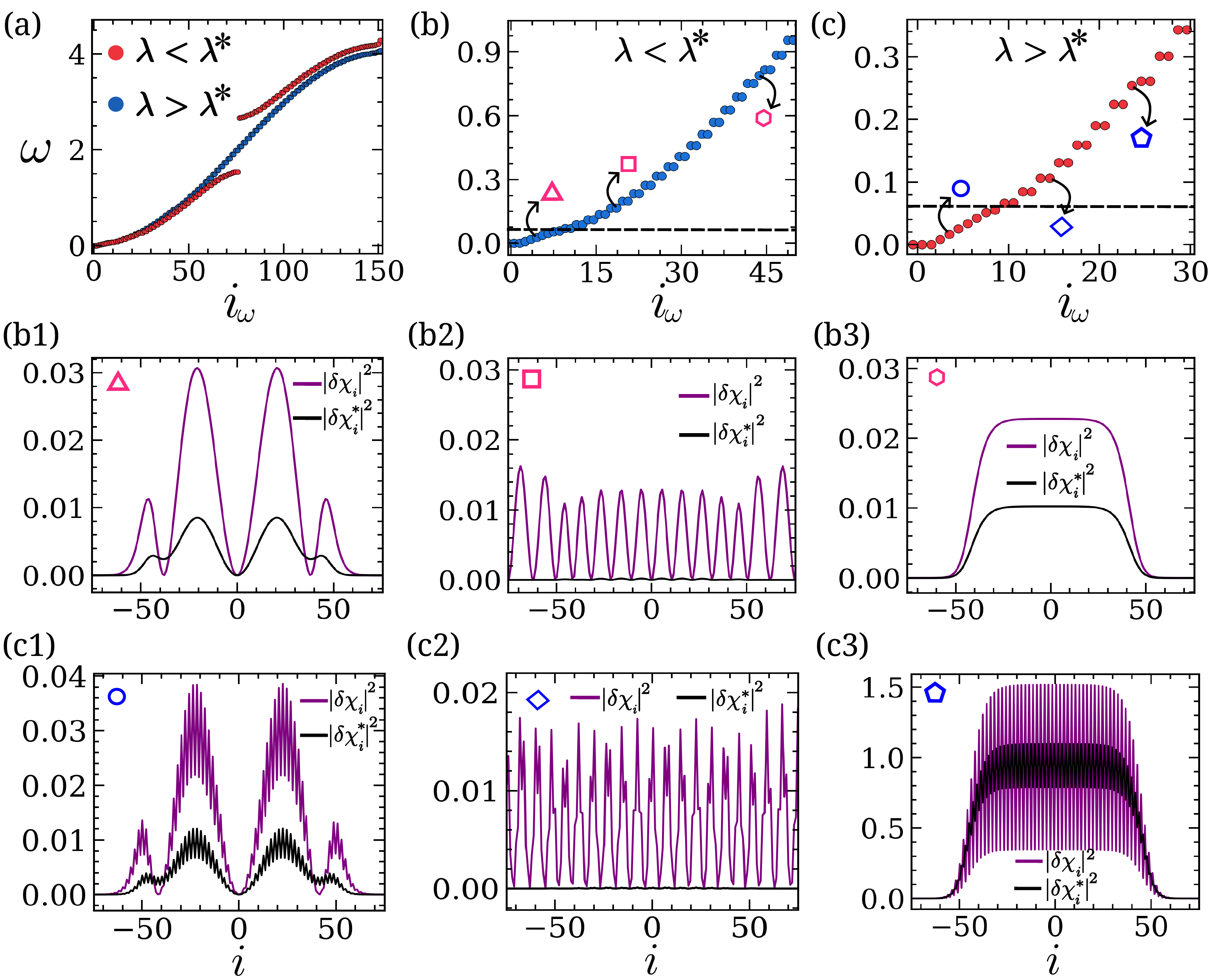}
	\caption{\textit{Excitation spectrum of the droplet states ($\zeta\neq0$)}: (a) Full excitation spectrum of the droplet phase before (blue) and after (red) the transition. Zoomed view of the low energy excitations of the (b) flat-top droplet case ($\lambda=0.5 < \lambda^{*}$) and (c) the staggered droplet phase ($\lambda=0.85 > \lambda^{*}$) are shown, where the dashed line ($-\mu$) denotes the threshold for the bound states. The markers in (b,c) correspond to the particle-like ($\delta \chi_{i}$, solid magenta lines) and hole-like ($\delta \chi^{*}_{i}$, solid black lines) fluctuations of (b1,c1) a localized bound state, (b2,c2) an extended scattering state, and the (b3,c3) collective polariton-like mode, for the flat-top and the staggered droplet phases, respectively. Parameters chosen are same as in Fig.~\ref{fig3}.}
	\label{fig4}
\end{figure*}

\begin{figure}
	\centering
	\includegraphics[width=0.7\columnwidth]{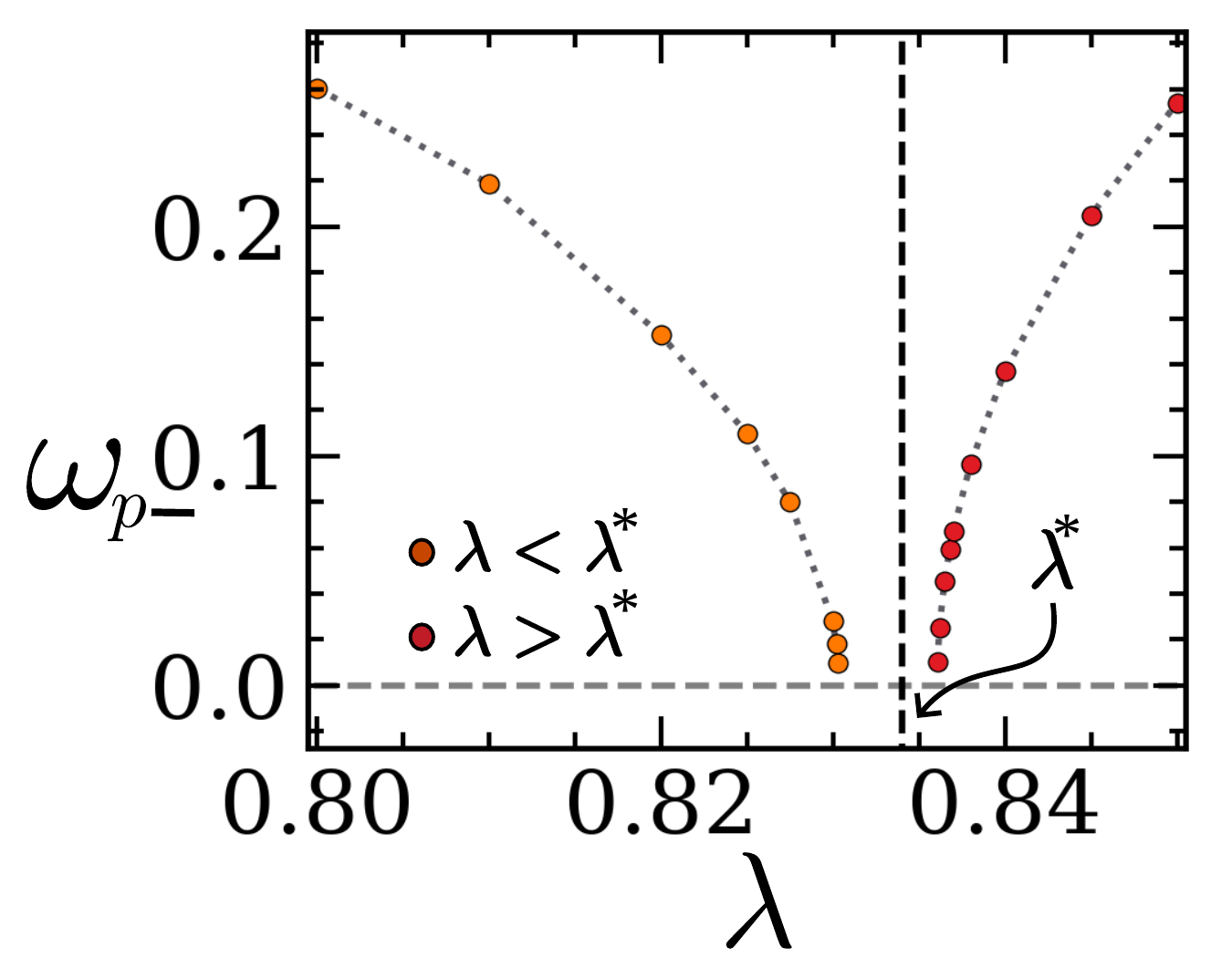}
	\caption{\textit{Polaritonic excitations of the droplet}:Variation of the collective polaritonic mode frequency $\omega_{p-}$ (lower branch) with coupling strength $\lambda$ across the crossover point $\lambda^*$. The behavior is similar to that observed for the homogeneous system in the presence of dissipation. The dotted lines only serve as a guide to the eye and the vertical dashed line marks the crossover point $\lambda^{*} \sim 0.834.$}
	\label{fig5}
\end{figure}

With increasing $\lambda$, the flat-top droplet undergoes a structural reorganization and crosses over to a density-modulated staggered configuration across the coupling strength $\lambda^*$, as shown in Fig.~\ref{fig3}(b). Above this crossover coupling, the total imbalance $\mathcal{I}$ associated with the density modulation increases with increasing coupling strength $\lambda$, and is also accompanied by the onset of a finite photon occupation density in the cavity [see Fig.~\ref{fig3}(c)]. Due to the emergence of superradiance in the cavity above the crossover coupling $\lambda^*$, we refer to the density-modulated state as the \textit{`superradiant droplet phase'}. Moreover, the non-vanishing cavity population serves as a direct probe of the amplitude of the sawtooth density modulation in the droplet. Note that, despite a finite light-matter interaction, below the crossover, the photon population associated with the flat-top droplet vanishes ($|\bar{\alpha}_d|^2=0$), as a consequence of the absence of density modulation.
Although the overall shape of the droplet remains same across the crossover, the average central density $\bar{n}_0^{\rm avg}$, obtained by averaging the two-sublattice densities, and the root mean square (RMS) width $w_d^{\rm RMS}=\sqrt{\sum_{j}j^2n_j/\sum_jn_j}$ (with the sums restricted between the edges of the droplet), exhibit small changes. As depicted in Fig.~\ref{fig3}(d), both quantities remain unaffected for $\lambda<\lambda^*$, whereas for $\lambda>\lambda^*$, $\bar{n}_0^{\rm avg}$ decreases slightly from $\bar{n}_0$, the central density of the uncoupled flat-top droplet [see Eq.~\eqref{eq:droplet_sat}], accompanied by a small increase in $w_{d}^{\rm RMS}$. Remarkably, the self-bound structure of the droplet is retained even for strong light-matter coupling and does not split in spite of the large density modulations.

Although the droplet density depends implicitly on the decay rate $\kappa$ through the effective coupling $\lambda_{\rm eff}$ in the energy functional [Eq.~\eqref{eff_energy}], weak dissipation does not qualitatively alter the overall shape of the droplet. This is consistent with our consideration of photon loss as the only source of dissipation, whose primary effect is to modify the cavity field and, consequently, the amplitude of the density modulation above the crossover. We find this behavior to persist even for moderate values of $\kappa$. It is possible that including other sources of dissipation, such as particle loss, may directly alter the shape of the droplet due to the dissipative effects on the density dynamics. For instance, in 2D exciton-polariton systems, such dissipation can significantly modify the condensate density due to the onset of the azimuthal currents \cite{Keeling2008_PRL}. We now turn to the excitation spectrum of the droplet and examine how dissipation influences the collective modes.

\textit{Excitations of the droplet:} Once the droplet phase is obtained as a non-equilibrium steady state in the lattice through the coupling to the cavity mode, its stability can be analyzed using small amplitude fluctuations, as mentioned in Sec.~\ref{subs:stability_analysis}. After achieving the stationary state, we construct the stability matrix given in Eq.~\eqref{eq:generalBdG} numerically and diagonalize it. The eigenvalues ($\{\Lambda_{i}\}$) obtained in turn determine the stability of the droplet for ${\rm Re}(\Lambda_{i})\leq0$ and yield the excitation frequencies $\omega_{i} = |{\rm Im}(\Lambda_{i})|$. Unlike the homogeneous system, the momentum is not a good quantum number to describe the excitations due to the spatially localized structure of the droplet and therefore we sort the excitations in ascending order according to their frequency for both the normal and superradiant droplet phases, as shown in Figs.~\ref{fig4}(a-c). Due to the localized structure of the droplet, the excitation spectrum can be divided into two parts separated by the energy threshold $\simeq -\mu$. Above this, the excitations form free particle-like continuum states which appear as doubly degenerate pairs due to the parity symmetry. Consequently, the particle-like fluctuations corresponding to $\delta \chi_{i}$ exhibits oscillatory and extended nature while the hole-like contribution associated with $\delta \chi_{i}^{*}$ vanishes, as evident from Figs.~\ref{fig4}(b2,c2). Moreover, the eigenvalues corresponding to these modes are completely imaginary and are decoupled from the photon fluctuations. On the other hand, the excitations below the threshold are non-degenerate and describe the bound modes of the droplet. For these excitations, both the particle and hole-like fluctuations, $\delta \chi_{i}$ and $\delta \chi_{i}^{*}$ respectively, form a localized structure, as depicted in Figs.~\ref{fig4}(b1,c1). The frequency of these bound modes correspond to the collective oscillations of the droplet such as the breathing mode, which play a very important role in defining the characteristic properties of the droplets \cite{Petrov2015, Astrakharchik2018, Malomed2020}. Furthermore, the number of bound modes reduces with decreasing the droplet size \cite{Malomed2020}.
Interestingly, sufficiently away from the crossover point $\lambda^{*}$, we identify two non-degenerate modes within the scattering states, for which both the particle and hole-like fluctuations are non-vanishing and localized, resembling the shape of the droplet [see Figs.~\ref{fig4}(b3,c3)]. These modes can be recognized as the polaritonic excitations due to the mixing between the condensate and photon fluctuations. Notably, these modes are dissipative in nature due to the negative real part of the corresponding eigenvalues, whereas for all other modes, the eigenvalues are purely imaginary. The variation of the lowest frequency $(\omega_{p-})$ of these polariton modes with $\lambda$ is shown in Fig.~\ref{fig5} which also softens in the vicinity of the crossover point similar to the homogeneous system. Although the droplet shape remains unaffected below the crossover coupling $\lambda^*$, its excitations, particularly the polaritonic modes, acquire a finite damping rate and can be probed dynamically by inducing small photon fluctuations around the corresponding steady state, as discussed later in Sec.~\ref{sec:real time and quench dynamics}.

\section{Metastable states: Kink and double-droplet solutions}
\label{sec:metastable solutions}

\begin{figure}
\centering
\includegraphics[width=\columnwidth]{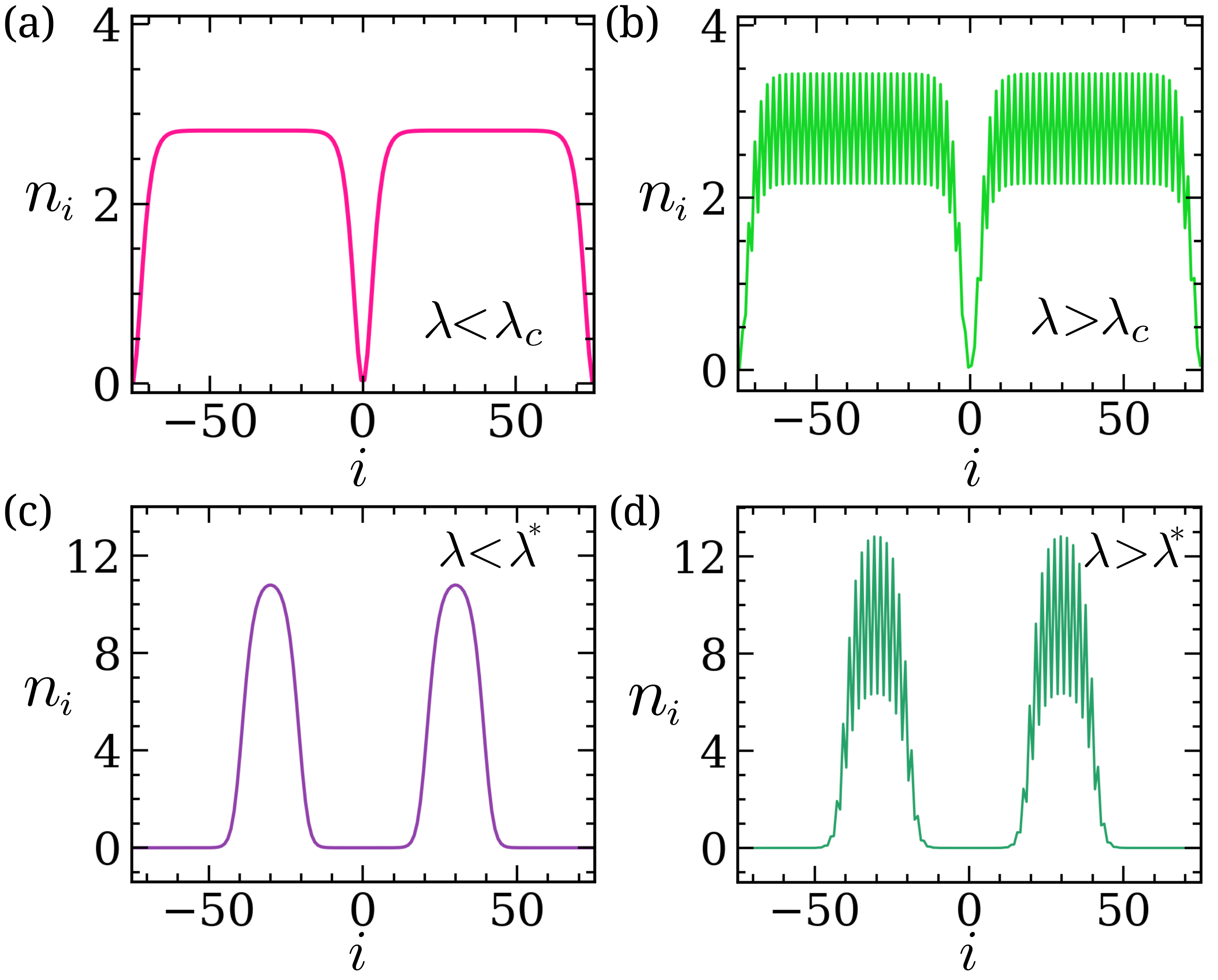}
\caption{\textit{Non-equilibrium (metastable) solutions}: Density profiles of the kink-like solutions (a)[(c)] below and (b)[(d)] above the transition [crossover] point $\lambda_{c}= 0.65 $ [$ \lambda^{*} \simeq 0.648$]. The condensate exhibits localized density dips (kinks) embedded in either a homogeneous or droplet background. Panels (a,b) correspond to the homogeneous case ($\zeta=0$); (a) for $\lambda = 0.2 < \lambda_{c}$, the density vanishes at the center and at the lattice boundaries but remains otherwise uniform, and (b) for $\lambda=0.66 > \lambda_{c}$, the background develops a staggered density modulation while preserving the kink structure. Panels (c,d) show the double-droplet solutions for $\zeta=0.2$; (c) for $\lambda=0.2 < \lambda^{*}$, the kink is embedded in a flat-top droplet background, and (d) for $\lambda=0.66 > \lambda^{*}$, the double-droplet background develops a staggered density modulation, corresponding to the superradiant double-droplet (SDD) state. Parameters chosen: $n_0=2.5$, $u=0.04$, and $\kappa=0.1$.}
\label{fig6}
\end{figure}
\begin{figure*}
	\centering
	\includegraphics[width=\textwidth]{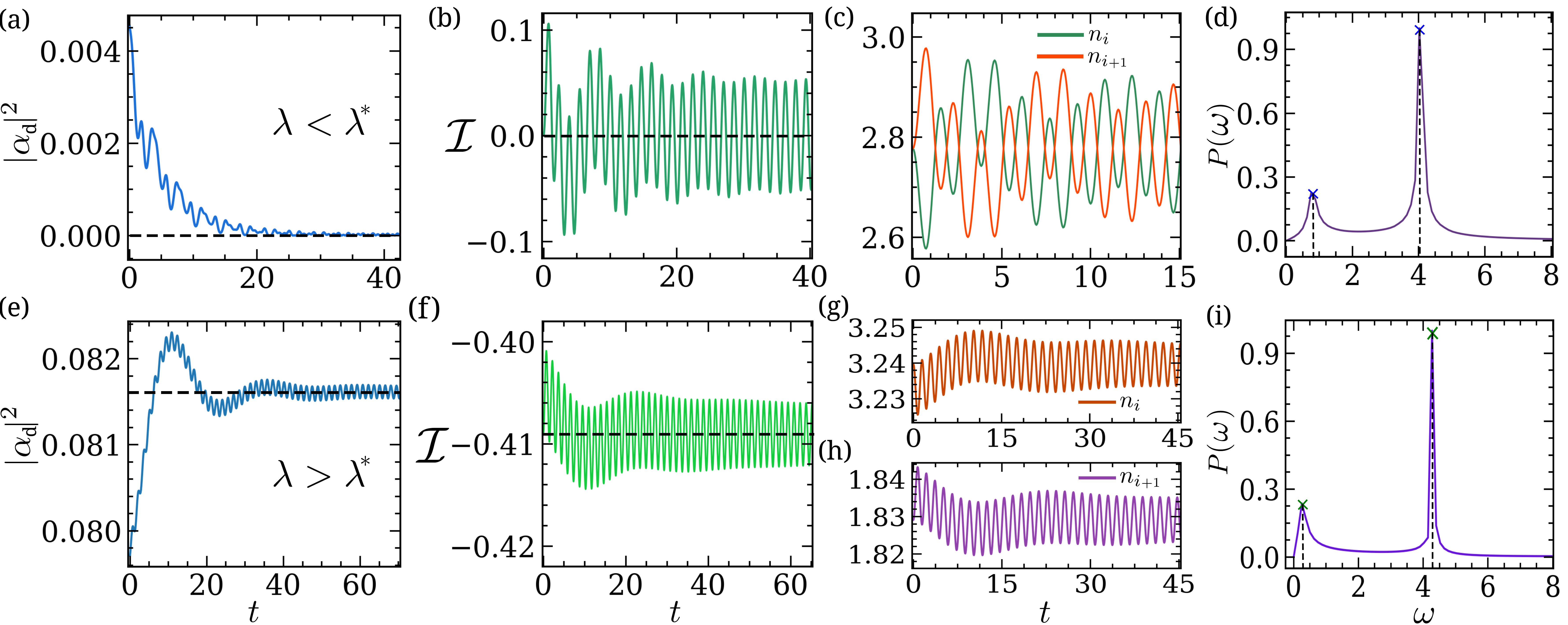}
	\caption{\textit{Real-time dynamics in the droplet regime}: The first row [panels (a)–(d)] corresponds to $\lambda= 0.5 < \lambda^{*}$, where the system evolves around the flat-top droplet steady state. The second row [panels (e)–(i)] correspond to $\lambda = 0.85 > \lambda^{*}$, where the evolution occurs around the staggered droplet state with finite photon number density. The crossover point is $\lambda^{*} \simeq 0.834$. Dynamics of the (a,e) photon number density $|\alpha_{d}|^2$ and (b,f) staggered imbalance $\mathcal{I}$. The black dashed lines denote the corresponding steady state values. (c,g,h) Dynamics of the local densities at neighboring sites, $n_{i}$ and $n_{i+1}$, exhibiting out-of-phase oscillations. (d,i) Fourier spectrum of $\mathcal{I}$ from (b,f). The dominant peaks coincide with the frequency of the collective polaritonic modes obtained from the stability analysis. Parameters chosen are same as in Fig.~\ref{fig3}.}
	\label{fig7}
\end{figure*}
So far, we have studied how cavity-mediated interactions induces a change in the non-equilibrium state of a homogeneous condensate and droplet phase into a superradiant density-wave state, through a dissipative transition. As mentioned previously, these steady states of the bosonic sector correspond to the ground state of the effective Hamiltonian [Eq.~\eqref{eff_energy}] obtained by eliminating the photon field. Consequently, these condensate wavefunctions have even parity symmetry. Here we investigate the other possible non-equilibrium states of the condensate having wavefunction with odd parity, since it is well known that the non-linear Schr\"{o}dinger equation permits a dark-soliton solution having a kink in the middle \cite{Kivshar1989, Shlyapnikov1999_PRL, Kivshar1994, Susanto2005, Pelinovsky2008, Yoshimura2019}. For this purpose, we minimize the effective energy functional [Eq.~\eqref{eff_energy}] within the odd-parity sector using imaginary-time evolution. Assuming a ring-like geometry with periodic boundary condition, we obtain a condensate configuration with two kinks, one at the center and the other at the boundary [see Fig.~\ref{fig6}(a)], satisfying both the odd parity and periodic boundary conditions. We further verify that the resulting condensate profile and chemical potential, obtained from imaginary-time evolution, satisfy the stationary conditions of the coupled equations [Eq.~\eqref{condensate_field},\eqref{cavity_field_dissipative}], with the corresponding photon number determined from Eq.~\eqref{SST2}.
Notably, a similar double-droplet like odd parity stationary state can also be obtained for the condensate with a competing attractive LHY interaction ($\zeta\neq0$), as shown in Fig.~\ref{fig6}(c). For sufficiently weak atom–photon coupling, both these solutions have a flat-top structure, while they develop a density wave modulation accompanied by a non vanishing photon field above the critical coupling strength, analogous to the homogeneous case [see Figs.~\ref{fig6}(b,d)]. Next we analyze the dynamical stability of the odd-parity solutions via linear stability analysis [Sec.~\ref{subs:stability_analysis}] by computing the eigenvalues of the stability matrix. Below a critical coupling, the states are stable for both $\zeta = 0$ and $\zeta \neq 0$, while above it they exhibit weak instabilities characterized by small positive real parts in the excitation spectrum. The effect of this weak instability can be further understood from the non-equilibrium dynamics and is discussed in the next section.

\section{Non-Equilibrium Dynamics and Collective Oscillations}
\label{sec:real time and quench dynamics}
After analyzing the steady states and their collective excitations, as well as the associated dissipative transitions, we now study the non-equilibrium dynamics of the photon coupled condensate, in the droplet regime. We focus on how the polaritonic excitations associated with the non-equilibrium transition can be probed dynamically and investigate the fate of the double-droplet state under a sudden quench across the crossover coupling.

\textit{Dynamical signatures of the polaritonic excitations} : The collective excitations of a finite condensate are usually measured by dynamically exciting certain collective coordinates \cite{Pitaevskii2016_book}. As discussed in Sec.~\ref{sec:droplet}, only two polariton-like collective modes arise due to the coupling between condensate and photon fluctuations. These modes can be selectively excited during dynamics by inducing photon fluctuations in the cavity. For this purpose, we consider the initial condensate corresponding to the steady state while choosing the photon field slightly away from its steady value and evolve it following Eqs.~\eqref{condensate_field} and \eqref{cavity_field_dissipative}. For below and above the transition, the collective mode of the condensate corresponding to the staggered density imbalance $\mathcal{I}$ oscillates with a slow decay, while the photon number quickly saturates to its corresponding steady value, see Figs.~\ref{fig7}(a,b),(e,f) respectively. The Fourier spectrum of the staggered density imbalance $\mathcal{I}$ shows two dominant peaks at frequencies corresponding to the two polaritonic modes of the system [see Figs.~\ref{fig7}(d,i)]. The lower-frequency mode is predominantly photon-like and rapidly relaxes, while the higher-frequency mode corresponds to condensate oscillations following the transient dynamics.
This is also evident from the relative contributions of the atomic and photonic components in the corresponding eigenvectors of the stability matrix. For the lower polaritonic branch, the atomic component of the associated eigenvector is negligible compared to its photonic component while the upper polaritonic branch exhibits a comparatively larger contribution from the atomic degrees of freedom.
In the bulk of the droplet, the densities at the neighboring central lattice sites show an out of phase oscillation around the corresponding steady state density as shown in Figs.~\ref{fig7}(c,g,h). For a homogeneous condensate, this mode generates an out of phase oscillation of the neighboring lattice sites around their corresponding steady values, exhibiting a spatio-temporal crystalline order. Even in the absence of any steady population imbalance, this mode generates a dynamics which oscillates between the two density wave patterns. Above the critical point, due to the explicit breaking of $\mathbb{Z}_2$ symmetry, the mode generates oscillations around one of the density wave configurations, corresponding to the steady state, as discussed in Appendix.~\ref{appendixC}.
\begin{figure}
	\centering
	\includegraphics[width=\columnwidth]{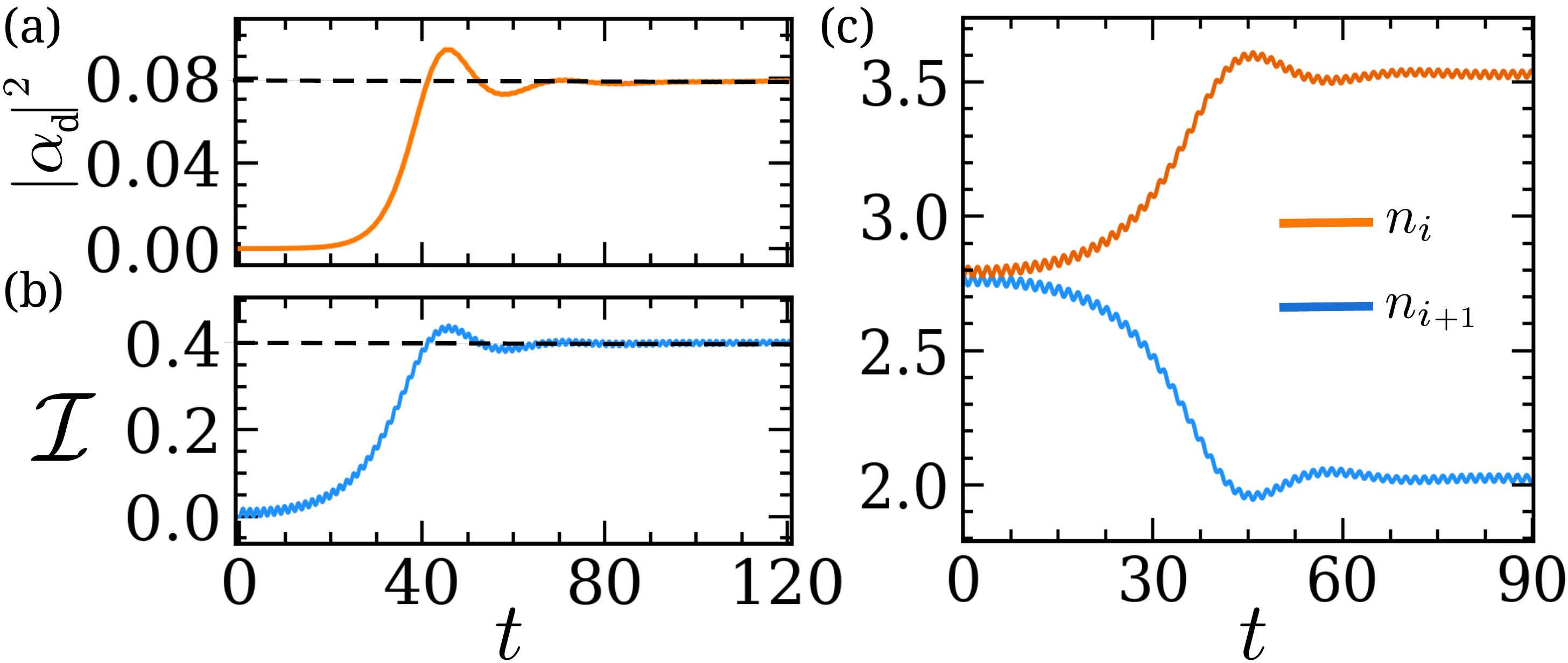}
	\caption{\textit{Quench dynamics across the crossover point starting from the flat-top droplet}: Time evolution of (a) photon number density $|\alpha_{d}|^2$ and (b) staggered imbalance $\mathcal{I}$, following a sudden quench of the atom-photon coupling strength from $\lambda=0.5$ to $\lambda=0.85$, starting from a flat-top droplet state. The crossover point is $\lambda^{*} \simeq 0.834$. In both cases, the black dashed line corresponds to the staggered steady state. (c) Dynamics of local densities at neighboring sites, exhibiting small oscillations around the sublattice densities. Parameters chosen are same as in Fig.~\ref{fig3} and $i=L/2$.}
	\label{fig8}
\end{figure}

\begin{figure}
	\centering
	\includegraphics[width=\columnwidth]{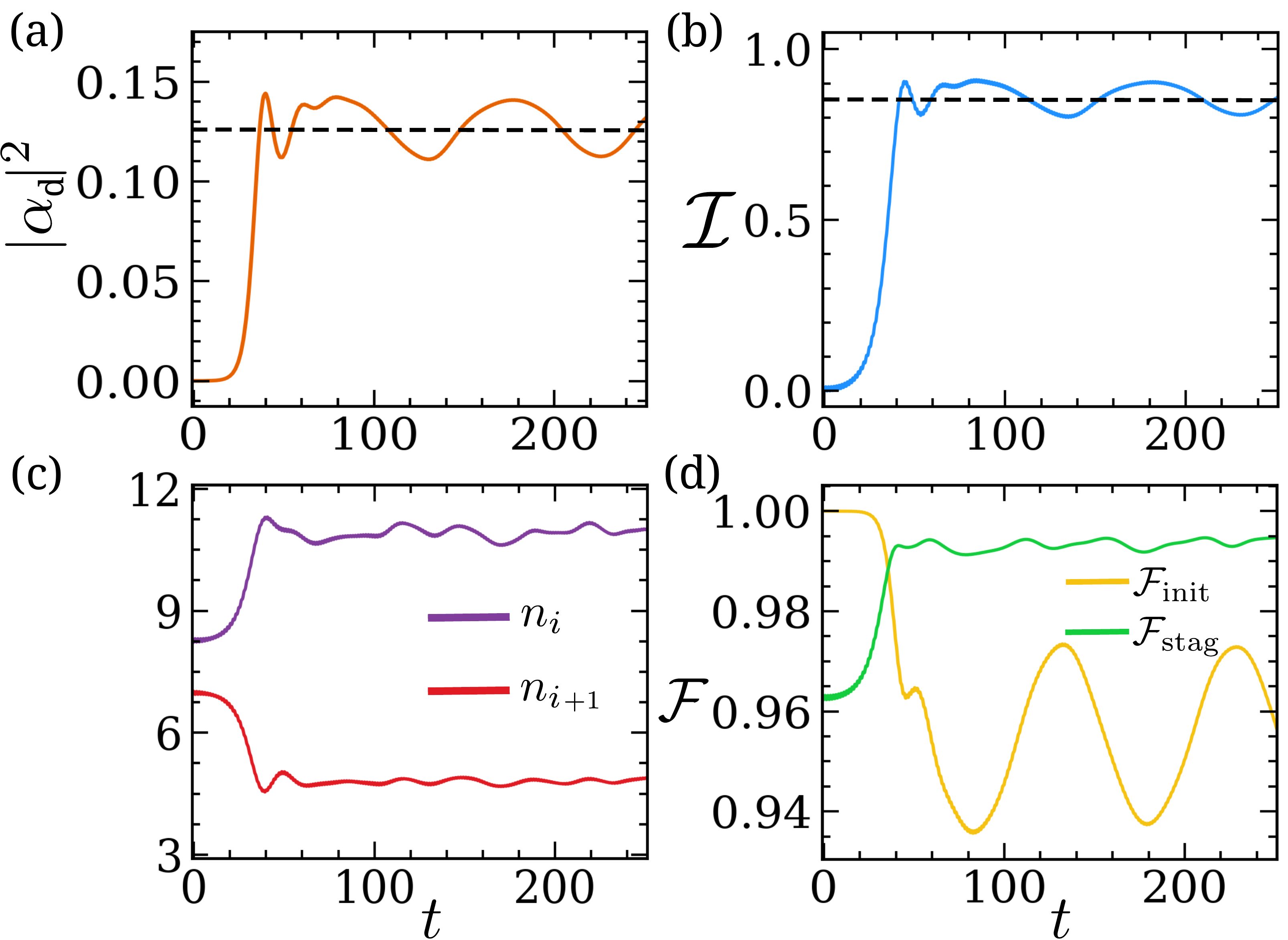}
	\caption{\textit{Quench dynamics across the crossover point starting from the flat-top double-droplet state}: Time evolution of (a) photon number density $|\alpha_{d}|^2$ and (b) staggered imbalance $\mathcal{I}$, following a sudden quench of the coupling strength from $\lambda=0.2$ to $\lambda=0.66$,  with the system initially prepared in the flat double-droplet configuration. The crossover point is $\lambda^{*} \simeq 0.65$. (c) Local densities at neighboring sites, $n_{i}$ and $n_{i+1}$, displaying oscillations around two distinct mean values associated with the emergent staggered background.(d) Dynamics of the fidelity with the staggered double-droplet state (green solid line) together with the survival probability relative to the initial flat-top double-droplet state. Parameters chosen:  $n_{0} = 2.5$, $u = 0.04$, $\zeta = 0.2$, and $i = L/4$.}
	\label{fig9}
\end{figure}

\textit{Quench dynamics}: Next, we study the non-equilibrium dynamics of the steady droplet state under sudden quench across the crossover coupling strength $\lambda^*$. We consider the flat-top stationary state of the droplet with vanishing cavity field as an initial state, corresponding to the steady state for $\lambda<\lambda^*$ and evolve it according to the EOMs [Eqs.~\eqref{condensate_field},\eqref{cavity_field_dissipative}] for $\lambda>\lambda^*$. After the quench, both the staggered density imbalance $\mathcal{I}$ and the photon number density $|\alpha_{d}|^2$ quickly saturate to the steady state value corresponding to the superradiant droplet [see Figs.~\ref{fig8}(a,b)] and the droplet develops a saw-tooth like density modulation. The dynamical generation of the density imbalance is reflected from the evolution of the densities at the neighboring lattice sites located at the center of the droplet, as shown in Fig.~\ref{fig8}(c).
Finally, we investigate the fate of the double-droplet state following a sudden quench above the crossover coupling, where it becomes unstable. We initialize the system in the stable flat-top double-droplet state prepared at $\lambda<\lambda^{*}$, and subsequently evolve it using Eqs.~\eqref{condensate_field} and \eqref{cavity_field_dissipative} for $\lambda>\lambda^{*}$. Remarkably, the double-droplet structure persists over sufficiently long times despite its instability in this regime. Although the survival probability $\mathcal{F}_{\rm init}$ decreases initially after the quench, it remains close to unity [see Fig.~\ref{fig9}(d)].Moreover, the condensate develops a density-wave modulation along with a finite photon occupation [Figs.~\ref{fig9}(a,b)], indicating the emergence of the superradiant phase with density-wave like order, reminiscent of the unstable superradiant double-droplet state. We also compute the overlap of the evolved state with the unstable superradiant double-droplet state [see Fig.~\ref{fig9}(d)], given by $\mathcal{F}_{\rm stag}=|\langle \Phi_{\rm SDD}|\phi(t)\rangle|^2,$ where $|\Phi_{\rm SDD}\rangle$ denotes the unstable superradiant double-droplet state. As shown in Fig.~\ref{fig9}(d), the overlap $\mathcal{F}_{\rm stag}$ rapidly increases and remains close to unity, exceeding the corresponding survival probability. This demonstrates that the time-evolved condensate follows the unstable superradiant double-droplet state with odd parity rather than relaxing to the stable even-parity superradiant single-droplet configuration. Due to the odd parity of the initially prepared double-droplet state, the dynamics remains largely confined to the odd parity sector even though such odd parity steady states are unstable in this regime. This behavior is similar to the collective scarring phenomena where the system dynamically follows the unstable steady state \cite{Mondal2024_PRA, Richter2023_PRL, Pizzi2025_natcomm, Sudip2024_review}. A similar phenomenon is also observed in the homogeneous phase, where the superradiant kink-like solution is likewise unstable.

\section{Discussions and conclusion}
\label{sec:summary}
In this work, we investigated non-equilibrium density ordering transitions as well as dynamics of dilute bosons confined to a one-dimensional lattice and coupled to a lossy cavity mode. Within a classical-field framework, we analyzed both the homogeneous condensate and self-bound droplet states. In both cases, increasing the atom–cavity coupling beyond a critical value drives the emergence of a density-wave ordered phase accompanied by a finite population of cavity photons, signaling a non-equilibrium superradiant transition. The formation of density-modulated superradiant droplets constitutes a particularly intriguing outcome of the interplay between quantum fluctuations and cavity-mediated interactions. The stability and collective modes of these non-equilibrium states were characterized through small-amplitude fluctuations around them. Besides the conventional sound-like Bogoliubov modes, we identify an additional collective excitation associated with the spontaneous breaking of the $\mathbb{Z}_{2}$ symmetry and the onset of density ordering.
As a precursor of this transition, the lowest branch of the polaritonic mode softens. However, unlike the QPT in isolated systems, its frequency vanishes in the vicinity of the critical point, while the relaxation time diverges as $1/|\lambda-\lambda_c|$, reflecting the non-equilibrium nature of the transition.
Such polariton-like mode is strongly coupled to the photon fluctuations, as a result of which, the effective Hamiltonian obtained within the adiabatic elimination approach fails to capture both the dispersion of this gapped mode as well as the features of this non-equilibrium transition. Experimentally, these excitations can also be probed by modulating the cavity field which in turn leads to the spatio-temporal oscillatory dynamics of the condensate. Above the transition point, the condensate density develops a two-sublattice structure, resulting in a reduced Brillouin zone where the Bogoliubov spectrum splits into two branches and a gap opens at the zone boundary. In the droplet regime, we further identified bound modes, which serve as characteristic signature of the self-bound structure. Finally, we demonstrated the existence of kink-like non-equilibrium configurations of the condensate and double-droplet states. These structures remain dynamically stable below the critical atom–cavity coupling and develop weak instabilities above the transition. Nevertheless, the double-droplet configurations can persist over long times even following a sudden quench across the crossover atom-photon coupling, highlighting the robustness and rich metastable dynamics of cavity-coupled quantum droplets similar to collective scarring phenomena.

In this work, we analyzed a non-equilibrium light-matter interacting system using a classical-field approximation. While such quasi-condensate description is only valid in 1D within certain appropriately chosen parameter regime \cite{Shlyapnikov2004, Castin2004}, the quantum fluctuations are typically more pronounced in lower dimensions. Consequently, the non-equilibrium dynamics of such Bose-gas or droplets coupled to a dissipative cavity mode may require further investigation using advanced techniques to incorporate the quantum fluctuations \cite{Astrakharchik2020, Chen2021, Astrakharchik2024, Kollath2020, Kraus2022, Orso2025}. Nevertheless, the present analysis demonstrates the potential of exploring rich non-equilibrium phenomena by coupling the condensate and self-bound droplets with a dissipative cavity mode which can be readily accessed using current experiments. It is worth mentioning that most previous studies of the Bose-Hubbard model in a cavity are based on an effective Hamiltonian obtained by adiabatically eliminating the photon field. However, the dynamics of the photon fluctuations are neglected in such an approach, which fails to capture the polaritonic excitations obtained in the present work. Furthermore, the decay rate of these excitations exhibits critical slowing down, revealing the non-equilibrium nature of the transition, which is absent in the effective Hamiltonian description. By retaining the dynamical nature of the cavity field, we provide a more complete description of the open light-matter coupled system and its non-equilibrium dynamics. Although, in the present framework, we describe the bosonic part using classical-field approximation, which is only valid in the deep superfluid regime, this formalism can in principle be extended to strongly correlated phases, particularly the Mott-Insulator phase, where the bosonic degrees can be treated using  more advanced methods. Furthermore, it would be also fascinating to study the dissipative dynamics of such macroscopic states in the presence of a structured bath which can be engineered using multi-mode cavity setups \cite{Sun2025_natcomm, Cheng2024_PRA, Plenio2018_NJP}.
\appendix
\section{Derivation of the Discrete Energy Functional for quantum droplet}
\label{appendixA}
In this section, we derive the Energy functional for the condensate on a periodic lattice [see Eq.~\eqref{energy_functional_discrete}] within the tight-binding approximation. For a binary mixture of Bose gases in quasi-1D geometry with equal masses ($m$), intra-species interaction, and densities, the energy functional of an effective single component BEC near the point of mean-field instability is given by \cite{Astrakharchik2016},
\begin{eqnarray}
\mathcal{E}_{b}[\psi,\psi^*] &=& \int \!dx \Bigg(\frac{\hbar^2}{2m}\,|\partial_x\psi|^2 + V(x)|\psi|^2 \notag\\
&&+ \frac{\delta g}{2}|\psi|^4 - \frac{2\sqrt{2m}}{3\pi\hbar}\, g^{3/2}|\psi|^3\Bigg). \label{energy_function_gas}
\end{eqnarray}
In the above equation, $\psi(x)$ denotes the classical field corresponding to the condensate with total number of particles $N = \int |\psi(x)|^2 dx \!\gg\! 1$, $V(x)$ is the external potential, $g = 2a_{s}\hbar^2/ml^2_{\perp} \!>\! 0$ represents the effective 1D interaction strength with $a_{s}$ being the intra-species s-wave scattering length, and $\delta g = g_{12} + g$ is the mean-field interaction strength of the effective single component where $g_{12}$ corresponds to the inter-species interaction strength. In general, if $g_{12}$ is attractive, then one can tune the parameters in such a way so that $\delta g>0$ as well as $\delta g/g \ll 1$, which corresponds to the point of mean-field instability. Under such circumstances, the mean-field interaction (second term) in Eq.~\eqref{energy_function_gas} becomes comparable with the beyond mean-field correction (third term), which is essential for the formation of quantum droplet \cite{Petrov2015,Astrakharchik2016}.

Considering the external potential to be periodic, $V(x) = V_{0}\sin^2{(2\pi x/\lambda_{L})}$, with $V_{0}$ ($\lambda_{L}$) being the potential depth (wavelength) of the optical lattice, we now apply the tight binding approximation, $\psi(x) = \sum_{j}\phi_{j}w(x-jd)$, where $d=\lambda_{L}/2$ is the lattice spacing, $|\phi_j|^2$ is the filling, and $w(x)$ is the Wannier state. Assuming that the lattice is deep enough i.e. $V_{0} \gg E_{R} = \hbar^2 \pi^2/2md^2$, we consider a Gaussian wavefunction of the form, $w(x) = (1/\pi \sigma^2)^{\frac{1}{4}}\exp(-x^2/2\sigma^2)$, where $\sigma \sim (d/\pi)(E_{R}/V_{0})^{\frac{1}{4}}$ is the corresponding width. After evaluating the integrals, Eq.~\eqref{energy_function_gas} modifies to following,
\begin{eqnarray}
\mathcal{E}_{b}[\phi_{j}, \phi^{*}_{j}] &=& \sum_{j}-J(\phi^{*}_{j}\phi_{j+1}+\phi^{*}_{j+1}\phi_{j}-2\phi^{*}_{j}\phi_{j}) \\
&+&\frac{1}{2}\frac{\delta g}{\sigma} \frac{|\phi_{j}|^4}{(2\pi)^{\frac{1}{2}}}\!-\! \frac{2}{3}\frac{2|\phi_{j}|^3}{\sqrt{3}\pi^{\frac{5}{4}}}\!\left(\!\frac{g}{\sigma\hbar\omega_{\perp}}\!\!\right)^{\!\!\frac{3}{2}}\!\frac{\sigma}{l_{\perp}}\hbar\omega_{\perp},\notag
\end{eqnarray}
where $J \simeq (4E_{R}/\sqrt{\pi})(V_{0}/E_{R})^{\frac{3}{4}}e^{-2\sqrt{V_{0}/E_{R}}}$ is the tunneling amplitude for $V_{0} \!\gg\! E_{R}$ \cite{Zwerger2003}. By setting $u = (\delta g/\sigma\sqrt{2\pi})$ and $\zeta = (2/\sqrt{3}\pi^{\frac{5}{4}})(g/\sigma\hbar\omega_{\perp})^\frac{3}{2}(\sigma/l_{\perp})\hbar\omega_{\perp}$, we obtain the desired discrete energy functional where the interaction potential corresponds to Eq.~\eqref{lhy_pot}.

In the present work, we have chosen the number of atoms per site as $n_0\sim1.5-2.5$, which is in line with experimentally realized fillings of ultracold atoms loaded into an optical lattice inside a cavity \cite{Esslinger2016_Nat}. Moreover, to remain in the weakly interacting regime, the on-site interaction strength is chosen such that $u<J$, thereby justifying the use of the classical-field approximation. We also choose a relatively small cavity decay rate ($\kappa = 0.1J$), consistent with the weak-dissipation regime assumed in the Lindblad description. This choice is not intended to represent a specific experimental configuration, as our primary aim is to investigate how dissipation modifies the nature of the transition and the associated collective excitations. Nevertheless, the chosen value of $\kappa$ is comparable to those considered in cavity-QED experiments \cite{Rodriguez2016_NC}. Note that, the parameters considered here also ensure that the LHY term is attractive but tuning them can lead to a dimensional crossover \cite{Petrov2018, Gajda2018}, where such correction can become repulsive.

\section{Excitations of the effective energy functional}
\label{appendixB}
Here, we derive the excitation spectrum of the effective energy functional [Eq.~\eqref{eff_energy} in the main text], obtained by adiabatic elimination of the photon field, for a homogeneous condensate. Starting from the corresponding effective Hamiltonian, we derive the EOMs for the condensate and perform a linear stability analysis, following the procedure outlined in Sec.~\ref{subs:stability_analysis}. Transforming the thus obtained fluctuation equations from real space to momentum space, and following the same procedure as in sub-Sec.~\ref{sub:homogeneous_excitations}, the fluctuations can be written as,
\begin{eqnarray}
\dot{\imath}\delta\dot{\chi}_k&=&\big[2J(1-\cos(k))+2un_0-\mu\big]\delta\chi_k+un_0\delta\chi_{-k}^*\nonumber\\
&+&4\big[u-\lambda_{\mathrm{eff}}\big]\bar{\chi}\epsilon\delta\chi_{k-\pi}+2u\bar{\chi}\epsilon\delta\chi_{\pi-k}^*\nonumber\\
&-&2\lambda_{\mathrm{eff}}\big(\bar{\chi}\delta_{k,\pi}+\epsilon\delta_{k,0}\big)\Big[\bar{\chi}(\delta\chi_{\pi}+\delta\chi_{-\pi}^*)\nonumber\\
&&\hspace{3cm}+\epsilon(\delta\chi_{0}+\delta\chi_{0}^*)\Big].
\end{eqnarray}
The excitations corresponding to $k\neq0,\pi$ ($k\neq\pi$) for $\lambda>\lambda_c$ ($\lambda<\lambda_c$) are identical to those obtained from the coupled atom-photon equations, which are given in Eqs.~\eqref{eq:uni_continuum} and \eqref{stag_continuum}. Below the transition, the  fluctuation with $k=\pi$ ($\delta\chi_{\pi}$) gives only one mode which softens as we approach the transition, similar to the lower polaritonic branch $\omega_{p-}$ obtained from the coupled atom-photon equations. Above the transition, the mixing of modes with $\delta\chi_0$ and $\delta\chi_{\pi}$, gives rise to one gapless mode and another gapped excitation. The gapped excitations are given as,
\begin{eqnarray}
\omega&=&\sqrt{16J^2+8n_0J(u-2\lambda_{\rm eff})},\hspace{0.3cm} \lambda<\lambda_c\\
\omega^{2}&=&\frac{-c_2+\sqrt{c_2^{2}-4c_{0}}}{2},\hspace{1.5cm} \lambda>\lambda_c
\label{eq:effpolariton_above}
\end{eqnarray}
where $c_2=8J^2+4J(\epsilon^2-\bar{\chi}^2)\lambda_{\rm eff}-2(u-\lambda_{\rm eff})[u(\bar{\chi}^4-10\bar{\chi}^2\epsilon^2+\epsilon^4)+8\bar{\chi}^2\epsilon^2\lambda_{\rm{eff}}],$
and $c_{0}=\bigl[4J^{2}-(\bar{\chi}^{2}-\epsilon^{2})^{2}(u-\lambda_{\rm eff})^{2}\bigr]\bigl[4J^{2}+4J(\epsilon^{2}-\bar{\chi}^{2})\lambda_{\rm eff}-(u-\lambda_{\rm eff})\{u(\bar{\chi}^{4}-34\bar{\chi}^{2}\epsilon^{2}+\epsilon^{4})+(\bar{\chi}^{4}+14\bar{\chi}^{2}\epsilon^{2}+\epsilon^{4})\lambda_{\rm eff}\}\bigr]$.
As evident from the above equations, both the gapped excitations for both sides of the transition vanish at the critical coupling strength $\lambda_c$, as shown in Fig.~\ref{fig1}(a) of the main text, reflecting the typical behavior for a discrete $\mathbb{Z}_2$ symmetry breaking transition. Although the collective excitations of the different $k$-modes obtained from the effective equations match with those obtained from the coupled equations, the frequencies of the polaritonic branches differ significantly.

\section{Dynamics of the homogeneous phase}
\label{appendixC}
\begin{figure}[b]
\centering
\includegraphics[width=\columnwidth]{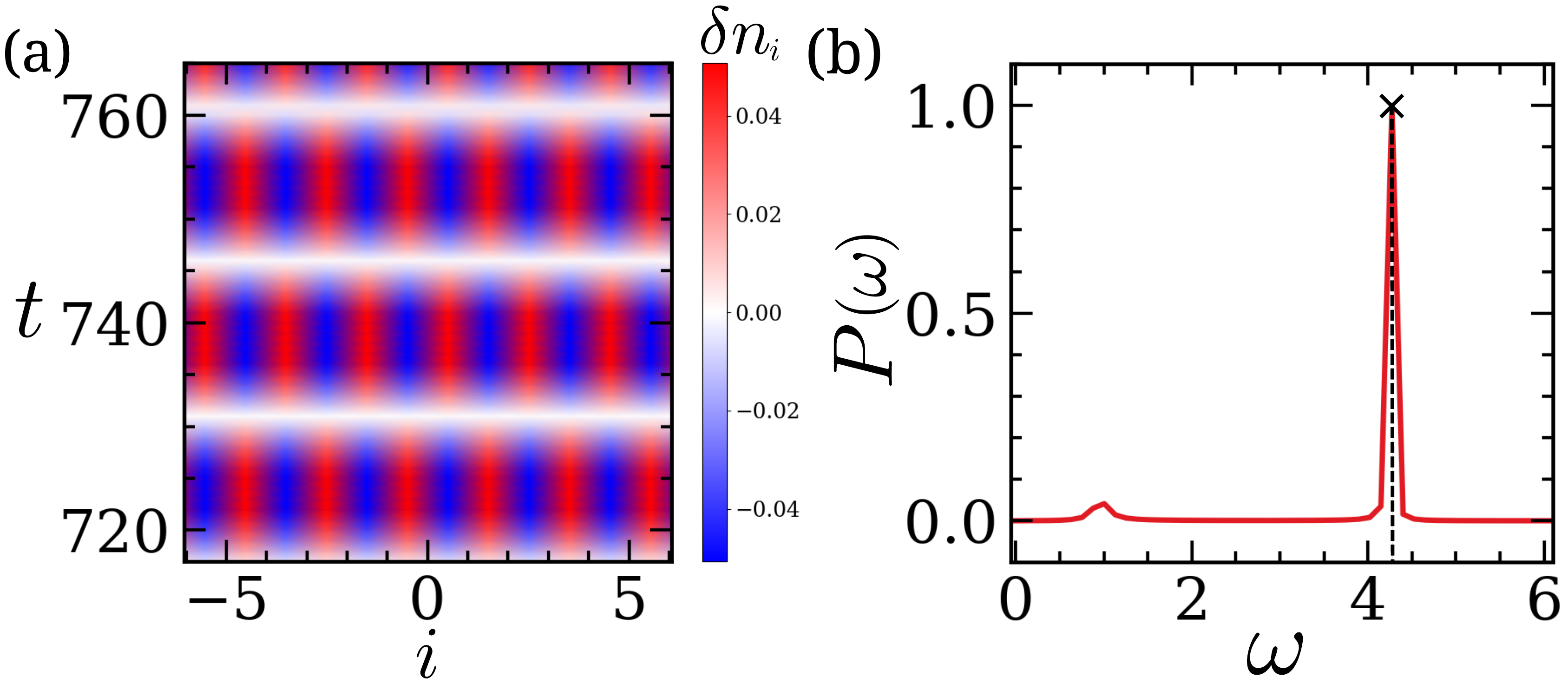}
\caption{\textit{Spatio-temporal crystalline structure in the homogeneous phase}: (a) Time evolution of the difference between the density at each lattice site and its steady state value, $\delta n_i=n_i-n_0$, shown as a color scale for $\lambda<\lambda_c$. Neighboring sites oscillate out of phase, resulting in a long-lived spatio-temporal density-wave pattern. For clarity, only the first few lattice sites are shown. (b) Fourier spectrum of $\delta n_i$ at the central lattice site ($i=L/2$). The dominant peak coincides with the polaritonic frequency $\omega_{p+}$ obtained from the linear stability analysis (see main text). Parameters: $n_0=2.5$, $u=0.1$, and $\kappa=0.1$.}
\label{fig_appC}
\end{figure}

To dynamically probe the polaritonic mode for a uniform condensate (with $\zeta=0$) at a coupling strength $\lambda<\lambda_c$,  Eqs.~\eqref{condensate_field} and \eqref{cavity_field_dissipative} are evolved, starting from a uniform condensate wavefunction corresponding to the steady state but with a finite fluctuation in the photon field. The evolution of the resulting density fluctuations $\delta n_i(t)=n_i(t)-n_0$ has been shown in Fig.~\ref{fig_appC}(a), after a transient time.
The perturbation generates a density modulation along the lattice moreover, the out of phase oscillation of the density fluctuations of the consecutive lattice sites gives rise to a spatio-temporal oscillatory behavior as shown in Fig.~\ref{fig_appC}(a). Although the steady state remains homogeneous with vanishing staggered density imbalance, the dynamics continuously oscillates between the two symmetry-related density-wave configurations. The Fourier spectrum corresponding to any lattice site shows a dominant peak corresponding to the polaritonic excitation with higher frequency $\omega_{p+}$, as shown in Fig.~\ref{fig_appC}(b). We can also observe a small contribution from the polaritonic excitation with lower frequency $\omega_{p-}$, however, the contribution of this mode is largely restricted to the short-time dynamics due to its larger decay rate.

\end{document}